\documentclass[twocolumn]{aastex631}

\usepackage{subfigure}
\usepackage{mathtools}
\usepackage{graphicx}
\usepackage{romannum}
\usepackage{comment}
\usepackage{textcomp}
\usepackage{booktabs}
\usepackage{multirow}
\usepackage{gensymb}
\usepackage{xcolor}
\usepackage{placeins}

\begin{document}
\title{A Study of Spectral Variability between flaring and non-flaring state in M74 X-1}

\author[0009-0000-2039-4340]{Aman Upadhyay}
\affiliation{Astronomy and Astrophysics, Raman Research Institute, C. V. Raman Avenue, Sadashivanagar, Bangalore 560080, India}

\author[0000-0002-3033-5843]{Tanuman Ghosh}
\affiliation{Inter-University Centre for Astronomy and Astrophysics, Ganeshkhind, Pune 411007, India}

\author[0000-0003-1703-8796]{Vikram Rana}
\affiliation{Astronomy and Astrophysics, Raman Research Institute, C. V. Raman Avenue, Sadashivanagar, Bangalore 560080, India}
 
\begin{abstract}
We conducted an extensive long-term spectral and timing study of the ultraluminous X-ray source M74 X-1, using data taken between 2001 and 2021 by Chandra and XMM-Newton X-ray observatories. Our analysis shows that flares are present in some observations, whereas they are absent in others. Flaring state exhibits two-component spectra at a lower average flux level, whereas non-flaring state displays single-component spectra at a higher average flux level. The M74 X-1 spectra are best described by the combination of accretion disk and Comptonization components, a dual thermal disk blackbody model, and a modified multi-temperature disk blackbody model. 
Using the dual thermal disk blackbody model, we obtain cool and hot temperatures of $T_{in}$ (cool) = $0.38^{+0.08}_{-0.06}$ keV and $T_{in}$ (hot) = $1.67^{+0.18}_{-0.13}$ keV, respectively, suggesting two temperature emitting regions and indicating possible presence of outflowing wind along with the accretion disk. We found a Gaussian feature at $E_{line}$ = $0.96^{+0.05}_{-0.11}$ keV with $\sigma$ = $0.11^{+0.13}_{-0.06}$ keV in the spectra of the flaring state which can be interpreted as the unresolved wind feature in the system when compared to similar feature seen in other ULXs. Plotting the hardness luminosity diagram, we get a trend of increasing hardness with luminosity, suggesting the presence of geometrical beaming in a low-inclination system. Additionally, using the hot disk blackbody component from the dual thermal disk blackbody model, we estimate the mass of the compact object to be M = $7.1^{+1.4}_{-1.3}$ M$_\odot$, classifying it as a stellar-mass black hole and confirming super-Eddington accretion in the system.

\end{abstract}


\section{Introduction} \label{sec:intro}
Ultraluminous X-ray sources (ULXs) are the extragalactic non-nuclear accreting binaries with X-ray luminosity exceeding the Eddington luminosity ($L_{X}$ $>$ 10$^{39}$ ergs/s) of a stellar-mass black hole ($\sim$ 10 $M_{\odot})$ \citep{king2023,kaaret2017}. Initially, these sources were perceived to be the scaled-up version of Galactic Black Hole Binaries (GBHBs) and were considered to be the Intermediate Mass Black Holes (IMBHs) accreting at sub-Eddington rates due to their low disk temperature \citep{colbert1999, miller2004}. However, the high-quality \textit{XMM-Newton} observations have shown that the ULXs spectra are characterized by features such as spectral curvature between 2-10 keV along with soft excess below 2 keV (e.g., \citealt{stobbart2006}), which are distinctive when compared to the GBHBs accreting at sub-Eddington rates. \cite{roberts2007} and \cite{gladstone2009} proposed that such spectral curvature is a characteristic feature of a new state, the \textit{ultraluminous state}, where the accretion is super-Eddington on a stellar mass compact object. This was further reaffirmed by the discovery of coherent pulsations in ULX M82 X-2, which proved the existence of a neutron star and hence the existence of super-Eddington accretion in the system (\citealt{bachetti2014}). Since then, several ULX pulsars have been discovered, which show pulsations \citep{israel2017a,israel2017b,furst2016,sathyaprakash2019,carpano2018, castillo2020} along with some ULXs which have shown pulsations in the outbursts (\citealt{tsygankov2017,wilson2018,vasilopoulos2020b}). Further, the discovery of Cyclotron Resonance Scattering Feature (CRSF) in M51 X-8 \citep{Brightman2018} revealed that even non-pulsating ULXs could harbor neutron stars.

\begin{table*}
\centering
\hspace{-2.0cm}
\begin{tabular}{c|c|c|c|c|c}
\toprule
Obs. ID & Mission & Detector& Flaring/Non-flaring & Date & ontime (ks) \\
\midrule
2057        & CHANDRA & ACIS-S & Flaring    & 19-06-2001 & 45.4 \\
2058        & CHANDRA & ACIS-S & Flaring    & 19-10-2001 & 46.1 \\
0154350101  & XMM-Newton & MOS1, MOS2, PN & Flaring     & 02-02-2002 & 28, 27, 22 \\
14801       & CHANDRA & ACIS-S & Non-flaring& 21-08-2013 & 9.8 \\
16000       & CHANDRA & ACIS-S & Non-flaring& 21-09-2013 & 39.6 \\
16001       & CHANDRA & ACIS-S & Non-flaring& 07-10-2013 & 14.7 \\
16484       & CHANDRA & ACIS-S & Non-flaring& 10-10-2013 & 14.7 \\
16485       & CHANDRA & ACIS-S & Non-flaring& 11-10-2013 & 8.9 \\
16002       & CHANDRA & ACIS-S & Non-flaring& 14-11-2013 & 37.6 \\
16003       & CHANDRA & ACIS-S & Non-flaring& 15-12-2013 & 40.4 \\
0864270101  & XMM-Newton & MOS1, MOS2, PN & Non-flaring & 13-01-2021 & 38, 39, 26 \\
\bottomrule
\end{tabular}
\caption{\protect\raggedright List of observations studied in this work taken by Chandra and XMM-Newton between 2001 and 2021.}
\label{tab:obs_set}
\end{table*}

Spectra of ULXs are broadly classified into three categories using a combination of an accretion disk and a power law model in the 0.3 - 10.0 keV energy range (\citealt{sutton2013}; \citealt{kaaret2017}). Typically, a broadened disk spectra appear like a single, broad continuum with inner disk temperature $T_{in}$ $>$ 0.5 keV and ratio of the flux of the power-law component to the flux of the disk component, $\frac{F_{PL}}{F_{disk}}$ {\small (0.3 - 1.0 keV) } $<$ 5. This spectral category corresponds to the lowest luminosity regime of ULXs ($\sim$ 1-3 $\times$ $10^{39}$ ergs/s). For $\frac{F_{PL}}{F_{disk}}$ {\small (0.3 - 1.0 keV) } $>$ 5, with $T_{in}$ $>$ 0.5 keV, the spectra are classified as an ultraluminous regime with a comparatively higher luminosity than Broadened disk state. For $T_{in}$ $<$ 0.5 keV, spectra are always in this ultraluminous regime. Depending on the power law photon index, this ultraluminous state is further classified into two different sub-categories, i.e., spectra with $\Gamma <$ 2 are categorized as ``Hard Ultra-Luminous" (HUL), whereas $\Gamma >$ 2 are classified as ``Soft Ultra-Luminous" (SUL) states. If the spectra are dominated by a single blackbody component with kT $\lesssim$ 0.1 keV and with the bolometric luminosity of $\sim$ $10^{39}$ ergs/s, then the source is defined as the Ultraluminous supersoft source\citep{kong2003,feng2016,pinto2017ultraluminous}.
\begin{figure*}
    \centering
    \includegraphics[width=\columnwidth]{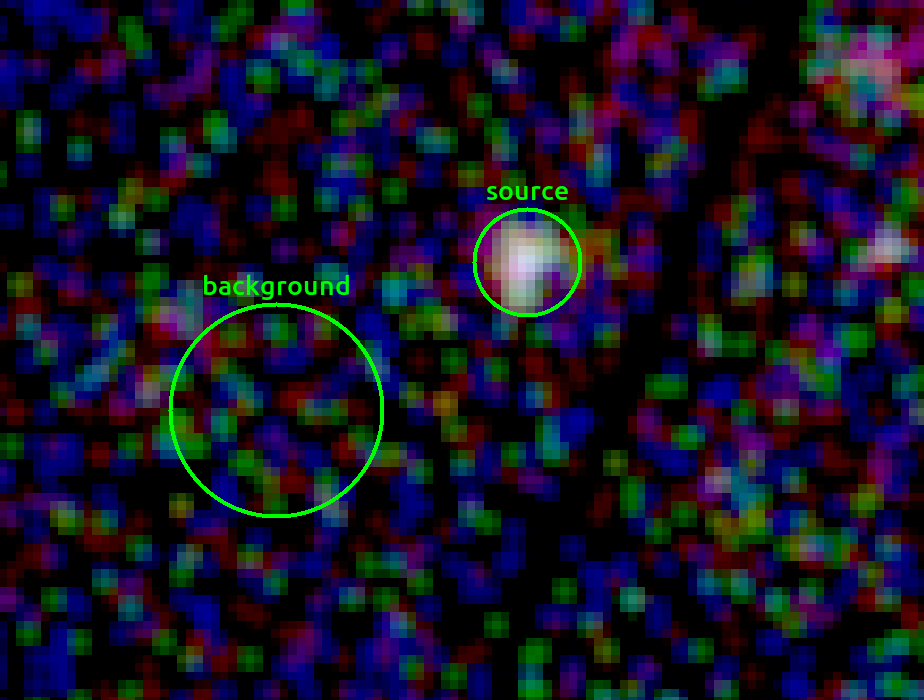}
         \label{fig:0154350101_MOS1}
    \includegraphics[width=\columnwidth]{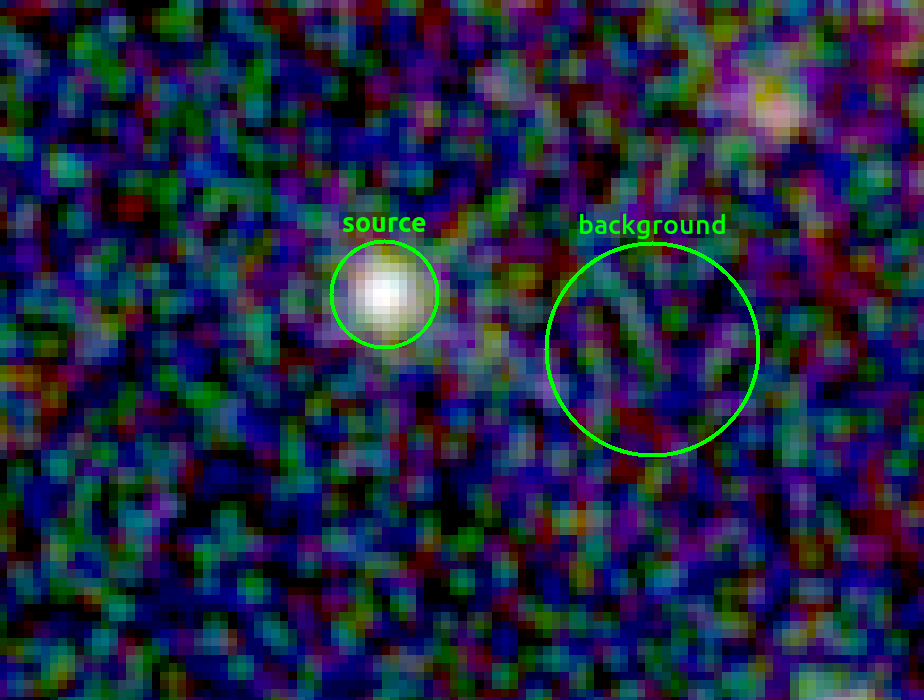} 
         \label{fig:0864270101_MOS1}
    \caption{\protect\raggedright Source and background region selection for Obs. ID 0154350101 (left) and 0864270101 (right) for the MOS1 detector. The figure is plotted in an RGB frame where Red corresponds to the events in (0.2 - 1.5 keV), Green corresponds to events in (1.5 - 2.5 keV), and blue corresponds to events in (2.5 - 10 keV).}
    \label{fig:xmm_images}
\end{figure*}
Broadband spectral analysis of ULXs has shown that their spectra in the 0.3-20 keV X-ray band can be characterized by two thermal disc components, while an additional component, such as coronal Comptonization or emission from the accretion column, is required to explain the high-energy band for non-magnetic and magnetic accretors, respectively (\citealt{walton2018, Walton2020}). The cool disk in the dual thermal disk model refers to the emission from the outer disk or the reprocessed soft photons coming from the optically thick outflows produced due to super-Eddington accretion in the system (\citealt{kajava2009,poutanen2007}), while the hot component comes from the inner regions of the accretion disk or corona (\citealt{walton2014,walton2015}; \citealt{luangtip2016}).\\ 

A large number of ULXs have been observed to date \citep{walton2022}, with some exhibiting persistent behavior while others showing extreme variability in which the flux can change over an order of magnitude. These variable sources have shown short-term timing variability in terms of periodic oscillation and quasi-periodic oscillation. Some of the sources like NGC 7456 ULX-1(\citealt{pintore2020}), NGC 6946 ULX-4 (\citealt{Earnshaw2019}), NGC 4559 X-7 (\citealt{pintore2021}), NGC 1313 X-1 (\citealt{Walton2020}), M51 ULX-7 (\citealt{earnshaw2016}), NGC 4395 ULX-1 (\citealt{ghosh2022}) and NGC 3621 ULX-1 (\citealt{motta2020}) have shown flaring events in the X-ray lightcurve which have helped in learning about the accretion processes in these sources. 
\par
M74 (NGC 628) is a nearby spiral galaxy located at a distance of 9.7 Mpc (\citealt{avdan2023}). It hosts two ULXs, M74 X-1 and M74 X-2, which are 1.5$\degree$ apart. Out of these, M74 X-2 is a transient source that was above the detection limit in only one XMM-Newton observation (Obs. ID: 0154350101) in February 2002 (\citealt{soria2002}). M74 X-1 has been studied earlier, and flaring activity was reported in the source with its luminosity varying in the range of $\sim$ 5 $\times$ $10^{38}$ ergs/s to $\sim$ 1.2 $\times$ $10^{40}$ ergs/s in a time period of half an hour (\citealt{Krauss2005}). 
\par
In this paper, we have performed an extensive long-term spectral and timing study of M74 X-1, using data from several observations taken between 2001 and 2021 by Chandra and XMM-Newton, listed in Table \ref{tab:obs_set}. Previously, \citealt{Krauss2005} and \citealt{LIU2005} studied 2001 Chandra and XMM observations and detected flaring events. Here we study all available observations from 2001 to 2021. The data reduction procedures are described in Section \ref{sec:data_red}. Results from spectral and timing analysis are reported in Section \ref{sec:analysis_results}. We discuss the physical implications of our analysis in Section \ref{sec:discussions} and conclude our findings in Section \ref{sec:conclusions}.

\section{Observation and Data Reduction}
\label{sec:data_red}
M74 was observed 16 times between 2001 and 2021 by XMM-Newton and Chandra. Of 16 observations, M74 X-1 was visible in 11 observations. In the remaining five observations, it was intrinsically faint, or the statistics were poor for any meaningful scientific analysis due to the low exposure time of the observations. We analyze all 11 observations (listed in Table \ref{tab:obs_set}), including the three previously studied observations in which flaring was reported (obs ID: 2057, 2058, 0154350101) by \cite{Krauss2005} and \cite{LIU2005}, and compare the spectral and timing properties between the flaring and non-flaring states.

\subsection{XMM-Newton}
The XMM-Newton data were extracted by using standard SAS software v21.0.0. The EPIC MOS and PN data were extracted using \texttt{emproc} and \texttt{epproc}. Significantly high background epochs were observed in both 2002 (Obs ID: 0154350101) and 2021 (Obs ID: 0864270101) observations, which were removed by creating good time intervals (GTIs) using the task \texttt{tabgtigen}. The durations of GTIs are 28, 27, 22 ks and 38, 39, and 26 ks for MOS1, MOS2, and PN for the 2002 and 2021 observations, respectively. These GTI files are used to create clean event files, which are further used to extract spectra and light curves using \texttt{evselect}. We apply the standard filter, i.e., {\small{PATTERN $<=$ 4}} for PN, and {\small{PATTERN $<=$ 12}} for MOS1 and MOS2. Barycentric correction of the events was done using the tool \texttt{barycen}. We use a $20''$ radius circle centered at R.A., Dec $=$ 138:40:33.74, -45:42:56.75 (\citealt{Krauss2005}) to extract source photons for both observations. We use a 40\arcsec radius region without source contamination from the same CCD to extract background photons for both observations (shown in Figure \ref{fig:xmm_images}). The spectra were binned using \texttt{specgroup} with a minimum of 20 counts per bin to allow for fits with $\chi^2$ statistics and an oversample of 3  to ensure that each group is at least 1/3 of full-width half-maximum resolution wide. Corresponding RMFs and ARFs were created using \texttt{rmfgen} and \texttt{arfgen}, respectively.
\begin{figure*}
    \centering

    \subfigure[Flaring lightcurve.
    \label{fig:flaring_lightcurve}]{
        \includegraphics[width=0.45\linewidth]{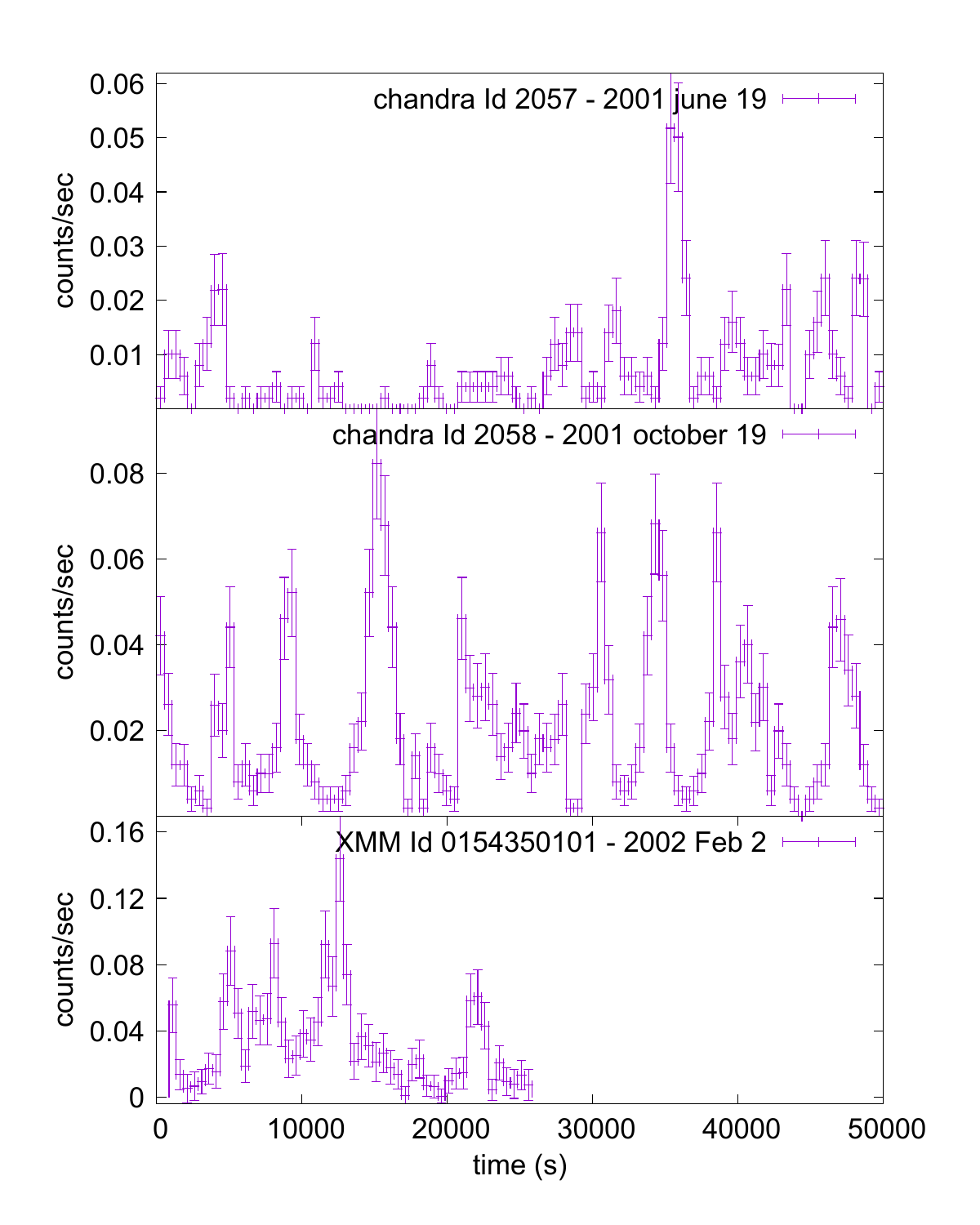}
    }
    \hfill
    \subfigure[Non-flaring lightcurve.
    \label{fig:non_flaring_lightcurve}]{
        \includegraphics[width=0.45\linewidth]{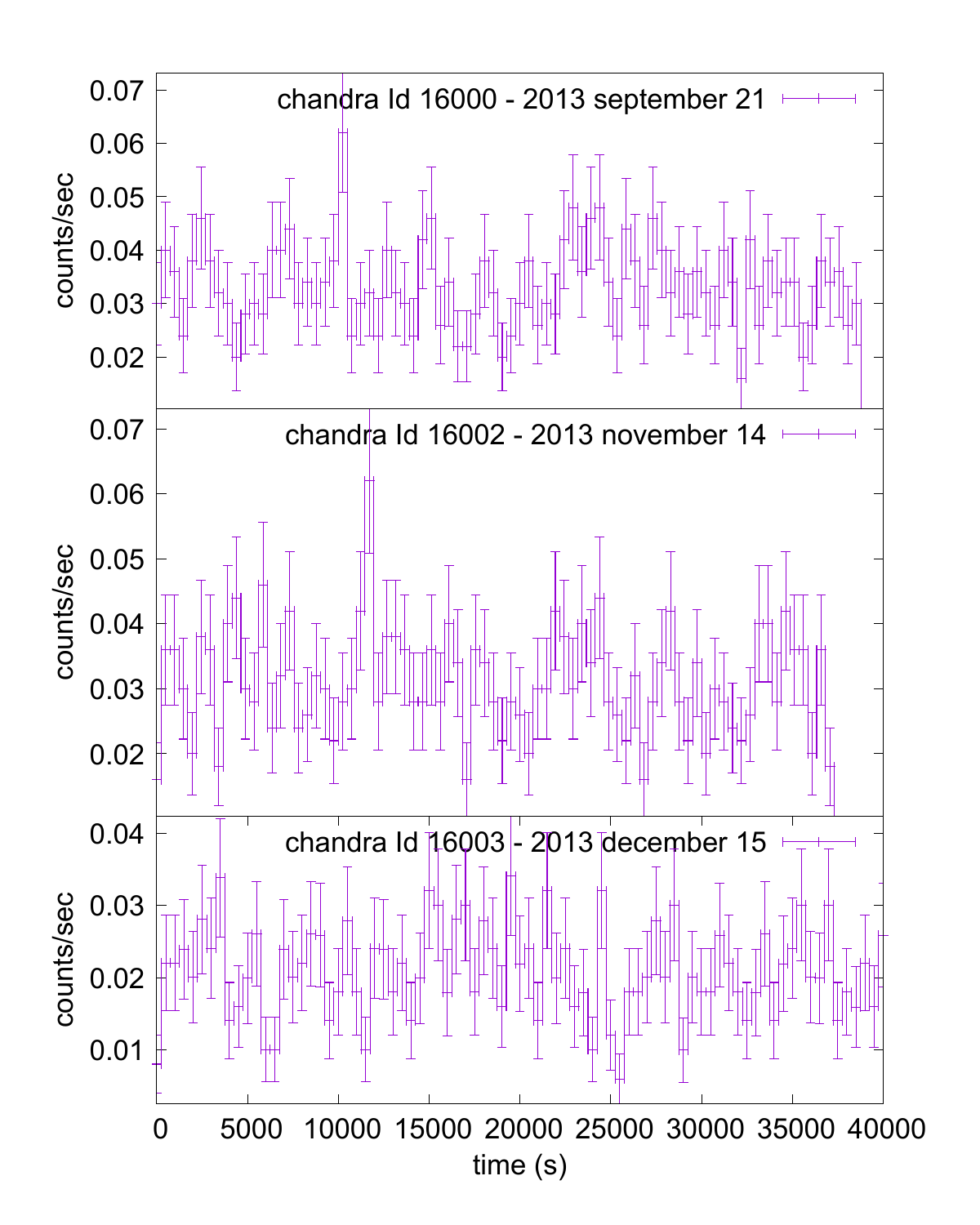}
    }

    \caption{Light curves of observations with flaring states (a) and non-flaring states (b), both binned with $t=500$ s. 
For the non-flaring state, lightcurves of observations (Obs. IDs: 16000, 16002, 16003) with the highest exposure time have been plotted.}
    \label{fig:lightcurve}

\end{figure*}
\begin{figure*}
    \centering
    \subfigure[Unfolded spectra of observations showing flaring activity. \label{fig:unfolded_spectra_flaring}]{
        \includegraphics[width=0.48\linewidth]{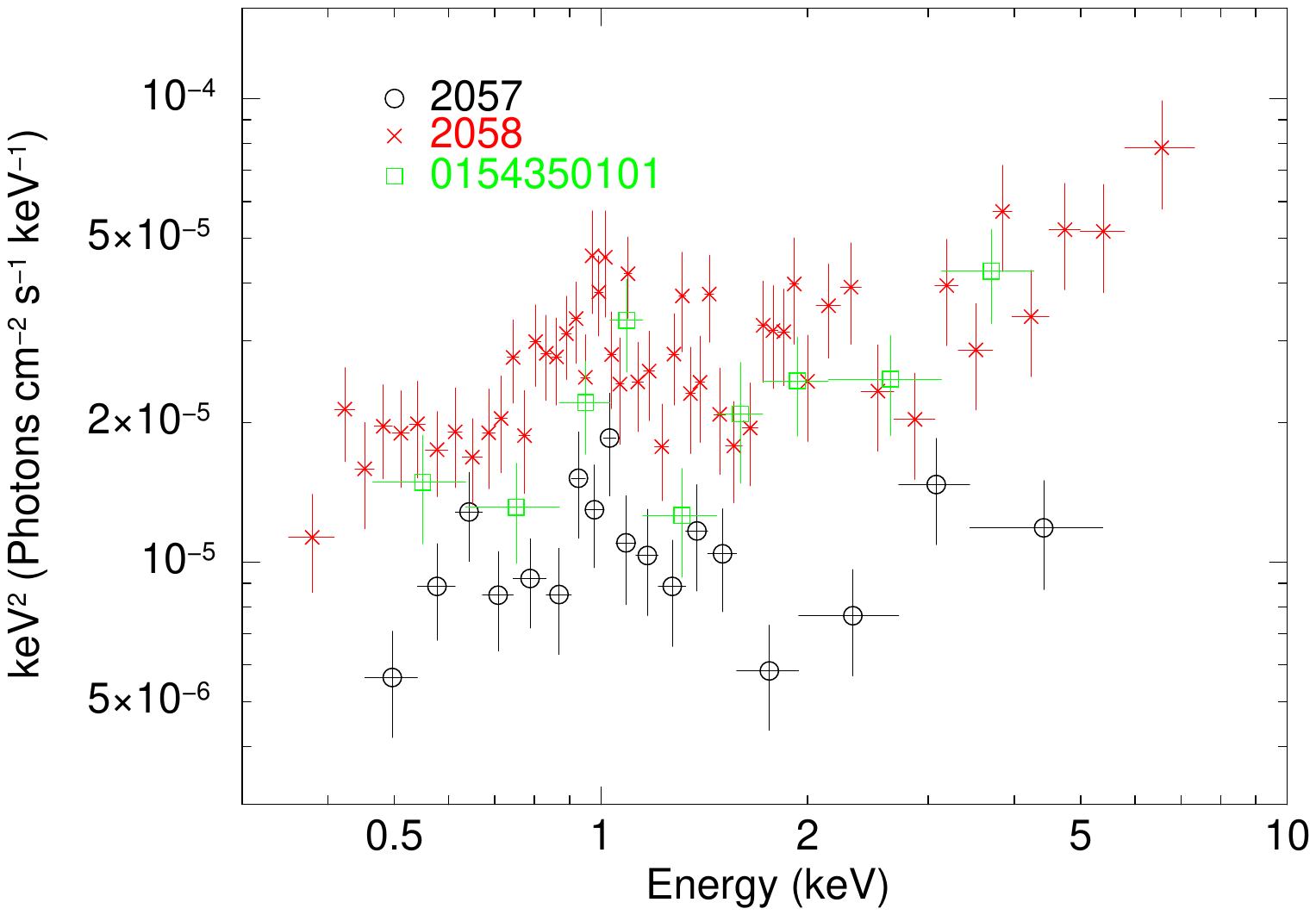}
    }
    \hfill
    \subfigure[Unfolded spectra of observations showing no flaring activity. \label{fig:unfolded_spectra_non_flaring}]{
        \includegraphics[width=0.48\linewidth]{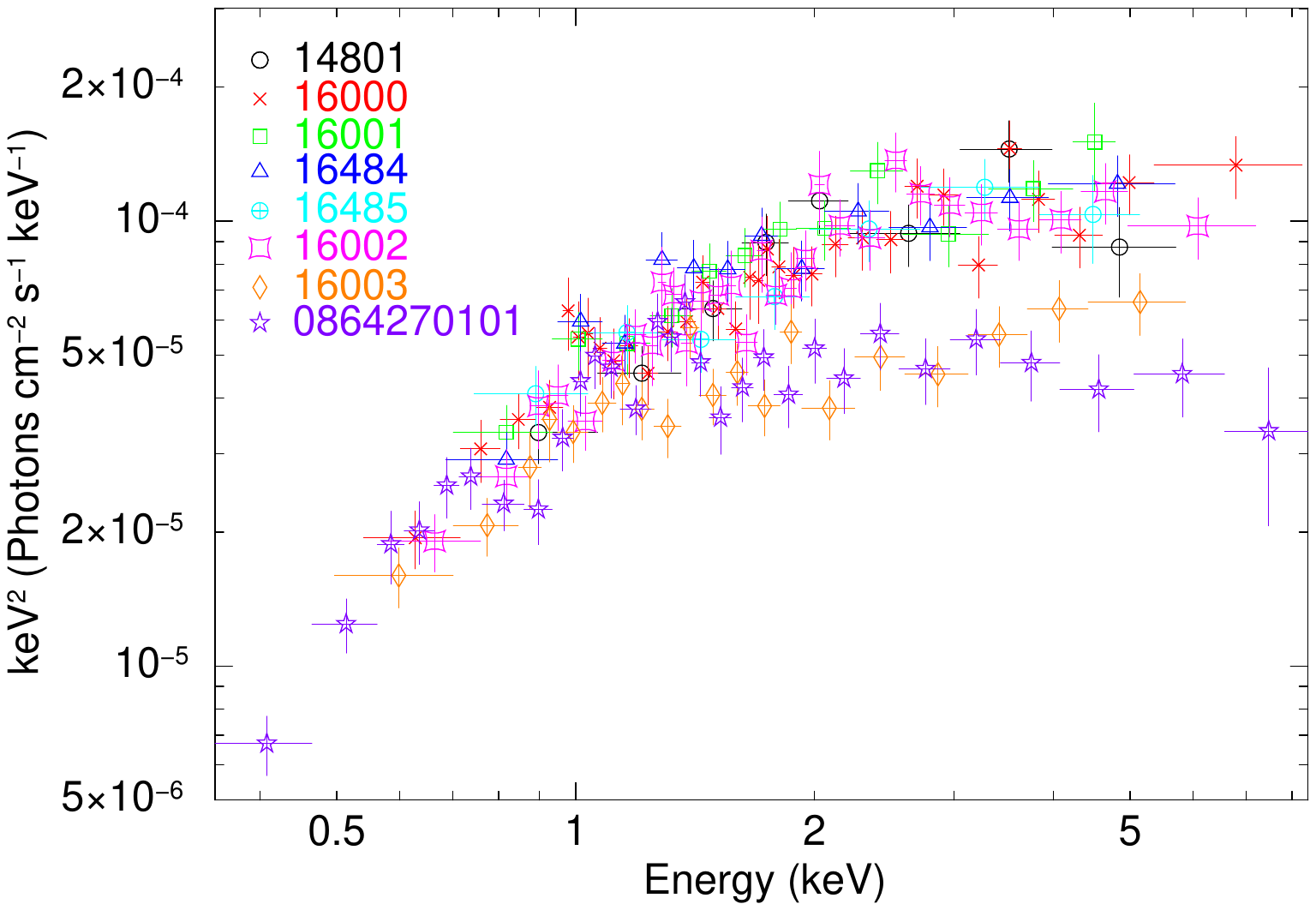}
    }

    \vspace{0.5cm}
    \subfigure[Difference in spectral shapes between spectra from Group A, B, and C. \label{fig:unfolded_spectra_groups}]{
        \includegraphics[width=0.5\linewidth]{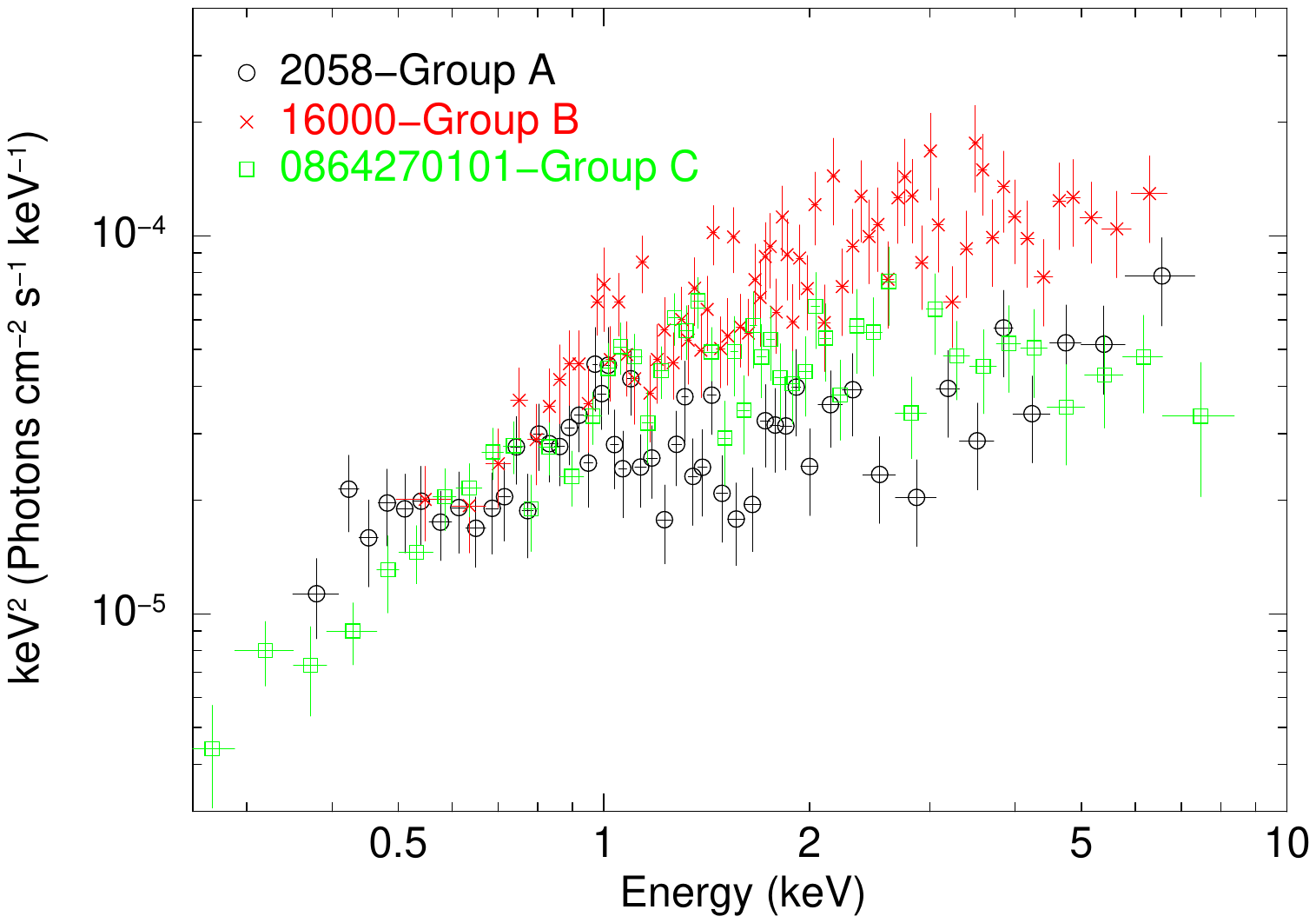}
    }

    \caption{Unfolded spectra of observations with flaring state (a) and observations with non-flaring state (b), using \texttt{powerlaw} model with zero photon index (constant model). Figure (a): All observations with flaring state have similar power law-like spectral shapes with different flux values. Figure (b): Spectra of observations with a non-flaring state seem to overlap below 1keV and are variable above 1keV. Figure (c): Different spectral shapes of the spectra from Group A, B, and C.}
    \label{fig:unfolded_spectra}
\end{figure*}

\subsection{Chandra}

We analyze Chandra ACIS-S data using Chandra Interactive Analysis of Observations (CIAO) v4.15 software, along with the calibration files CALDB v4.10.4. The level 2 event files are obtained with \texttt{chandra\_repro} in CIAO. \texttt{Wavdetect} is used to identify the source pixels in the image, which is further used to create a background file by excluding these pixels from the image. All observations are searched for the time intervals of high background flaring, and little to no background flaring was seen in all Chandra observations, which is also removed using the command \texttt{deflare}. To verify that the pileup does not affect the spectrum, we use the pileup algorithm implemented in PIMMS, where we find that the pile-up fraction in all the observations is $<5\%$. For source and background region selection, we use elliptical and circular regions for different observations according to the spread of the photons on the detector. Elliptical areas were selected for Obs. ID 2057, 2058, and 16003, whereas for the rest of the Chandra observations, circular source regions of radius 1.5\arcsec are selected, which covers the Full Point Spread Function (PSF) of the sources, and a $3.0''$ region on the same CCD is selected as background region. Further, \texttt{dmextract} and \texttt{specextract} are used to extract light curves and spectra of all the observations.

\section{Analysis and Results}
\label{sec:analysis_results}
\subsection{Timing analysis} 
Previous studies such as \citet{Krauss2005} have reported that M74 X-1 is a highly variable source as seen across observations (Obs. IDs - 2057, 2058, 0154350101; \citealt{Krauss2005}) with the count rate changing by an order of magnitude as illustrated in Figure \ref{fig:flaring_lightcurve}. We verify the flaring activity in these observations and find that the rest of the observations have a stable light curve with no flaring activity, as e.g., shown in Figure \ref{fig:non_flaring_lightcurve}. These flaring events are present in both hard and soft X-ray bands, which we verified by plotting light curves in the soft (0.3 - 1.0 keV) and hard (1.0 - 10.0 keV) X-ray energy bands (see Figure 9 in \citealt{Krauss2005}). Earlier studies (\citealt{Krauss2005}; \citealt{LIU2005}) have also suggested the existence of shallow frequency 2-hour quasi-periodic oscillations (QPOs) in the power density spectrum (PDS) of the observations with flaring state. Recently, \cite{avdan2024} has claimed a similar QPO feature in two more observations (Obs ID: 16485 and 16002) in a non-flaring state. However, the significance of these QPOs is considerably lower compared to the flaring state \citep{avdan2024}. To search for pulsation in the data, we employ an accelerated search technique with \texttt{HENACCELSEARCH} (a task in package HENDRICS v8.0.3; \citealt{bachetti2018}). We use this technique for both Chandra and XMM-Newton (EPIC-PN) observations in the energy range (0.3 - 10.0 keV), in the frequency range (0.01 - 6.8 Hz) and (0.01 - 0.15 Hz), for XMM-Newton and Chandra respectively, to avoid the artifacts due to the Nyquist limit. We do not see any sign of pulsations in these observations.
\begin{table*}[!htbp]
\scalebox{0.90}{
\hspace{-3.3cm}
\centering
\setlength{\tabcolsep}{6pt}
\renewcommand{\arraystretch}{1.5}
\begin{tabular}{c|c|c|c|c|c|c|c|c|c}
\toprule
\multicolumn{1}{c}{obs ID} & \multicolumn{1}{c}{nH} & \multicolumn{2}{c}{Power law} & \multicolumn{2}{c}{Disc Blackbody} & \multicolumn{3}{c}{Gaussian} & $\chi^2$/dof\\
\cmidrule(lr){1-1} \cmidrule(lr){2-2}  \cmidrule(lr){3-4} \cmidrule(lr){5-6} \cmidrule(lr){7-9} \cmidrule(lr){10-10}
 & $10^{22}(cm^{-2})$ & $\Gamma$ & Norm ($10^{-5})$ & Tin(keV) & Norm & $E_{l}(keV) $& $\sigma$(keV) & Norm ($10^{-5})$ \\
\midrule
\multirow{3}{*}{2057} & \multirow{9}{*}{$0.048^{\textcolor{red}{\textit{a}}}$} & $2.12^{+0.20}_{-0.19}$ & $1.01^{+0.10}_{-0.11}$  & -&-&-&-&-&$15/11$\\
&  & $\prescript{\textcolor{blue}{\textit{b}}}{}{1.29^{+0.30}_{-0.35}}$ &  $0.41^{+0.21}_{-0.17}$ &  $\prescript{\textcolor{violet}{\textit{d}}}{}{0.28^{+0.04}_{-0.04}}$ & $0.19^{+0.18}_{-0.08}$&-&-&-&$10/9$\\
&  & $\prescript{\textcolor{cyan}{\textit{c}}}{}{1.54^{+0.17}_{-0.26}}$ &  $0.62^{+0.16}_{-0.13}$ &  $\prescript{\textcolor{orange}{\textit{e}}}{}{0.15^{+0.12}_{-0.07}}$ & $>$0.13&$\prescript{\textcolor{green}{\textit{f}}}{}{0.96^{+0.05}_{-0.11}}$&$\prescript{\textcolor{magenta}{\textit{g}}}{}{0.11^{+0.13}_{-0.06}}$&$0.25^{+0.17}_{-0.17}$&10/6\\
\cline{1-1}
\cline{3-9}
\multirow{3}{*}{2058} &  & $1.88^{+0.10}_{-0.10}$ & $2.76^{+0.16}_{-0.16}$ & -&-&-&-&-&$44/39$\\
&  & $\prescript{\textcolor{blue}{\textit{b}}}{}{1.29^{+0.30}_{-0.35}}$ & $1.44^{+0.61}_{-0.54}$ &$\prescript{\textcolor{violet}{\textit{d}}}{}{0.28^{+0.04}_{-0.04}}$  & $0.40^{+0.31}_{-0.17}$&-&-&-&36/39\\
&  &   $\prescript{\textcolor{cyan}{\textit{c}}}{}{1.54^{+0.17}_{-0.26}}$ &  $2.00^{+0.35}_{-0.54}$ & $\prescript{\textcolor{orange}{\textit{e}}}{}{0.15^{+0.12}_{-0.07}}$ & $>$0.20&$\prescript{\textcolor{green}{\textit{f}}}{}{0.96^{+0.05}_{-0.11}}$&$\prescript{\textcolor{magenta}{\textit{g}}}{}{0.11^{+0.13}_{-0.06}}$&$0.45^{+0.28}_{-0.27}$&26/38\\
\cline{1-1}
\cline{3-9}
\multirow{3}{*}{01545350101} &  & $1.88^{+0.15}_{-0.15}$ & $2.06^{+0.29}_{-0.29}$  & -&-&-&-&-&$29/27$\\
&  & $\prescript{\textcolor{blue}{\textit{b}}}{}{1.29^{+0.30}_{-0.35}}$ & $1.11^{+0.51}_{-0.43}$ & $\prescript{\textcolor{violet}{\textit{d}}}{}{0.28^{+0.04}_{-0.04}}$ &$0.28^{+0.25}_{-0.13}$&-&-&-&27/27\\
&  & $\prescript{\textcolor{cyan}{\textit{c}}}{}{1.54^{+0.17}_{-0.26}}$ & $1.55^{+0.34}_{-0.43}$ & $\prescript{\textcolor{orange}{\textit{e}}}{}{0.15^{+0.12}_{-0.07}}$ &$>$0.10&$\prescript{\textcolor{green}{\textit{f}}}{}{0.96^{+0.05}_{-0.11}}$&$\prescript{\textcolor{magenta}{\textit{g}}}{}{0.11^{+0.13}_{-0.06}}$&$0.30^{+0.25}_{-0.23}$&20/26\\
\bottomrule
\end{tabular}
}
\caption{\protect\raggedright Spectral fitting results of observations in flaring state (Group A).\textcolor{red}{ \hspace{0.1cm}\textit{a} :} nH fixed to the galactic value.\textcolor{blue}{ \hspace{0.1cm}\textit{b}} and{ \textcolor{cyan}{\textit{c}}} : linked $\Gamma$ between all observations for model \texttt{tbabs*(diskbb+po)} and \texttt{tbabs*(diskbb+po+gauss)}. \textcolor{violet}{\hspace{0.1cm}\textit{d}} and \textcolor{orange}{\textit{e}}: linked $T_{in}$ between all observations for model \texttt{tbabs*(diskbb+po)} and \texttt{tbabs*(diskbb+po+gauss)}.\textcolor{green}{\hspace{0.1cm}\textit{f}} and \textcolor{magenta}{\textit{g}}: linked line energy $E_{l}$ and line width $\sigma$ respectively between all observations for \texttt{tbabs*(diskbb+gauss+po)}.
}
\label{tab:flaring_obs_fitting}
\end{table*}

\begin{table*}
\scalebox{1.0}{
\hspace{0.5cm}
\centering
\setlength{\tabcolsep}{6pt}
\renewcommand{\arraystretch}{1.5}
\begin{tabular}{c|c|c|c|c|c}
\toprule
\multicolumn{1}{c}{obs ID}&\multicolumn{1}{c}{nH}& \multicolumn{3}{c}{Cutoff power law} &$\chi^2$/dof\\
 \cmidrule(lr){1-1} \cmidrule(lr){2-2} \cmidrule(lr){3-5} \cmidrule(lr){6-6}
 & $10^{22}(cm^{-2})$ & $\Gamma$ & $E_{fold}$(keV) & Norm ($10^{-5})$&  \\
\midrule
14801 & \multirow{8}{*}{$\prescript{\textcolor{red}{\textit{a}}}{}{0.14^{+0.05}_{-0.04}}$}  & \multirow{6}{*}{$\prescript{\textcolor{blue}{\textit{b}}}{}{1.02^{+0.27}_{-0.27}}$} & \multirow{8}{*}{$\prescript{\textcolor{green}{\textit{d}}}{}{4.63^{+3.5}_{-1.4}}$} & $6.77^{+0.77}_{-0.73}$ & 15/11\\

16000 &  &   & &$6.70^{+0.50}_{-0.47}$&61/57\\

16001&  &   & &$7.56^{+0.72}_{-0.68}$&17/22\\

16484 &  &  & &$7.43^{+0.71}_{-0.67}$&20/23\\

16485 &  &  & &$6.68^{+0.79}_{-0.75}$&13/12\\

16002 &  &   & &$6.50^{+0.51}_{-0.48}$&57/50\\
\cline{3-3}
16003 &  &   \multirow{2}{*}{$\prescript{\textcolor{blue}{\textit{c}}}{}{1.52^{+0.28}_{-0.28}}$}  & &$5.30^{+0.48}_{-0.44}$&38/36\\

0864270101 &  &   & &$5.30^{+0.56}_{-0.52}$ & 106/103\\

\bottomrule
\end{tabular}
} 
\caption{\protect\raggedright Spectral fitting results of observations in non-flaring state (Group B + Group C) with \texttt{cutoff} power law model with linked nH and $E_{fold}$ and $\Gamma$. Simultaneous fitting with \texttt{tbabs*(cutoffpl)} gives $\chi^2$/dof = 327/313. \textcolor{red}{ \hspace{0.1cm}\textit{a}} and \textcolor{green}{ \hspace{0.1cm}\textit{d}} : nH and $E_{fold}$ values linked between all observations. \textcolor{blue}{ \hspace{0.1cm}\textit{b}} and \textcolor{blue}{ \textit{c}}: linked $\Gamma$ values in group B and group C respectively.}
\label{tab:non_flaring_cutoffpl}
\end{table*}
\begin{figure*}
    \centering
    \hspace{-1.5cm}
    \subfigure[Spectra and residuals from the simultaneous fitting of observations in flaring state 
    (Obs. IDs: 2057, 2058, 0154350101) with models \texttt{tbabs*(po)}, \texttt{tbabs*(diskbb+po)}, 
    \texttt{tbabs*(diskbb+gauss+po)} for linked $N_{\rm H}$, $T_{\rm in}$, $\Gamma$ and Gaussian parameters.
    \label{fig:flaring_po_diskbb_gauss}]{
        \includegraphics[width=0.60\linewidth, height=0.35\textheight]{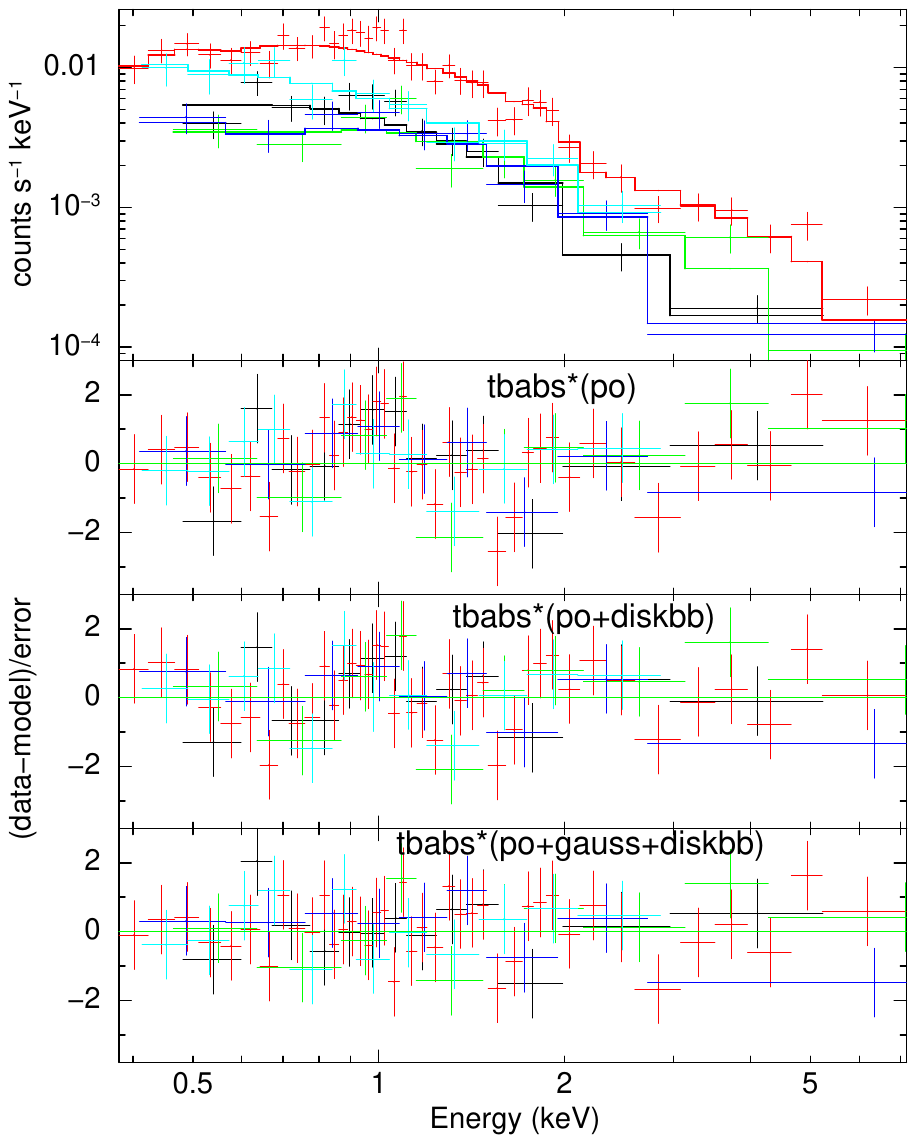}
    }
    \hfill
    \hspace{0.05cm}
    \subfigure[Spectra and residuals of observations in non-flaring state with model 
    \texttt{tbabs*(po)} with linked $N_{\rm H}$.
    \label{fig:non_flaring_spectra_res}]{
        \includegraphics[width=0.42\linewidth, height=0.25\textheight]{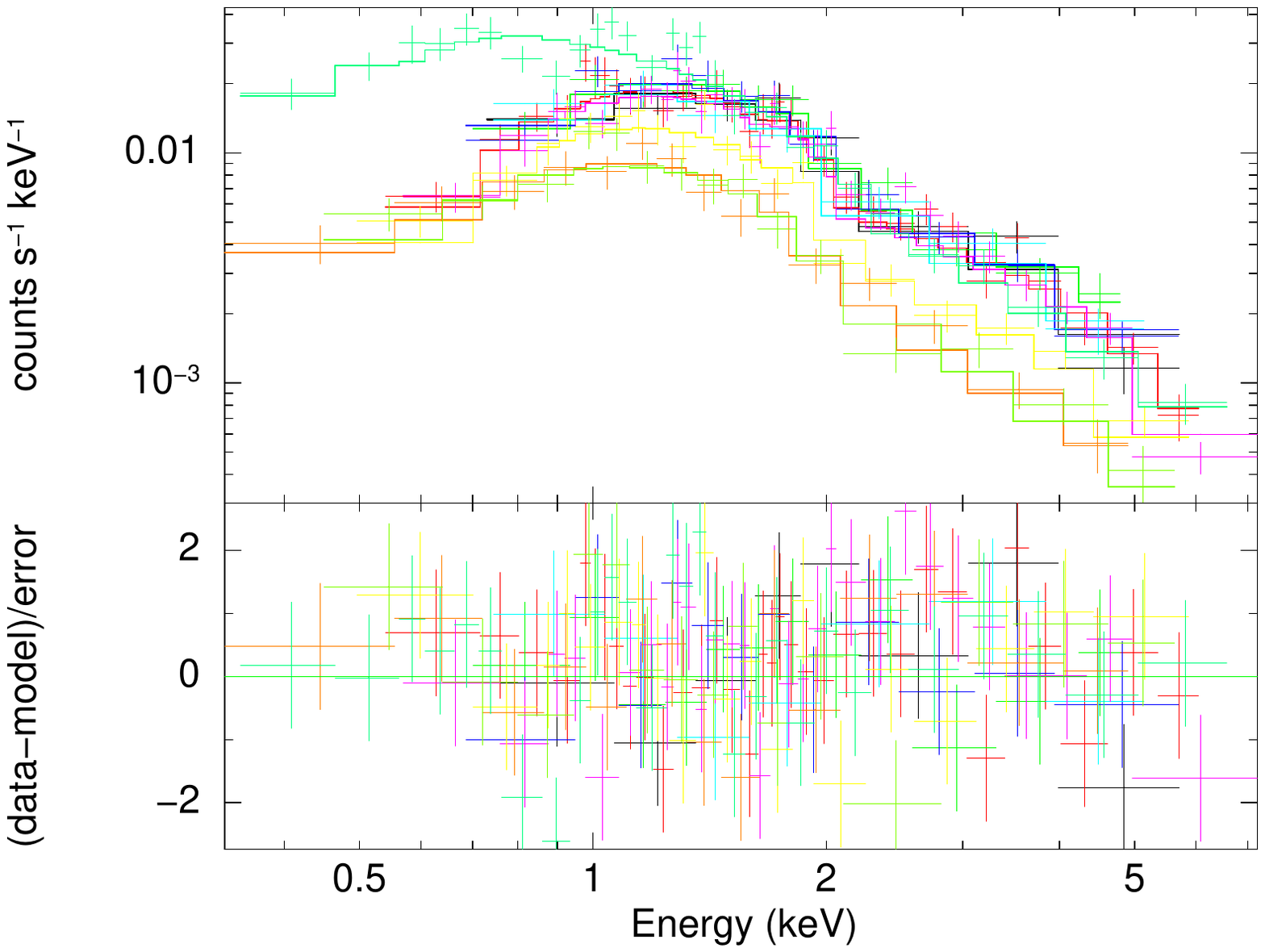}
    }

    \vspace{0.5cm}
    \subfigure[$\Gamma$ evolution of observations with non-flaring state using 
    \texttt{tbabs*(po)} with linked $N_{\rm H}$.
    \label{fig:gamma_evolution_non_flaring}]{
        \includegraphics[width=0.45\linewidth]{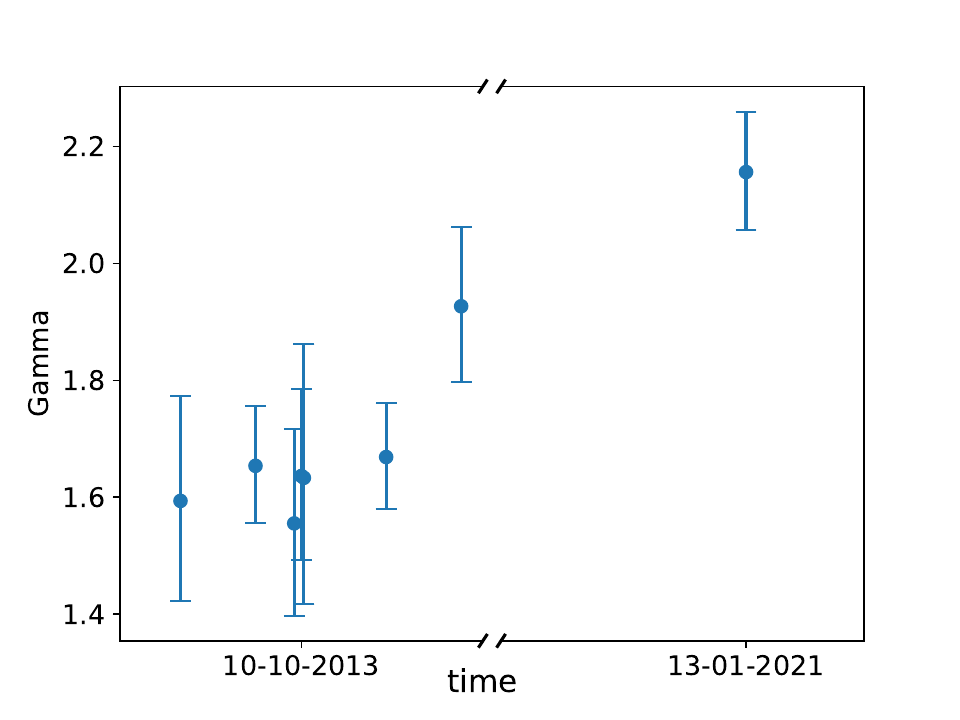}
    }
    \hfill
    \subfigure[Hardness–Luminosity diagram for all observations. Hardness is defined as the flux ratio 
    between (1.0–10.0 keV) and (0.3–1.0 keV), while luminosity is calculated in the 0.3–10 keV band. 
    Red circles represent flaring-state observations, blue squares represent non-flaring-state observations.
    \label{fig:Hardness_luminosity_diagram}]{
        \includegraphics[width=0.45\linewidth]{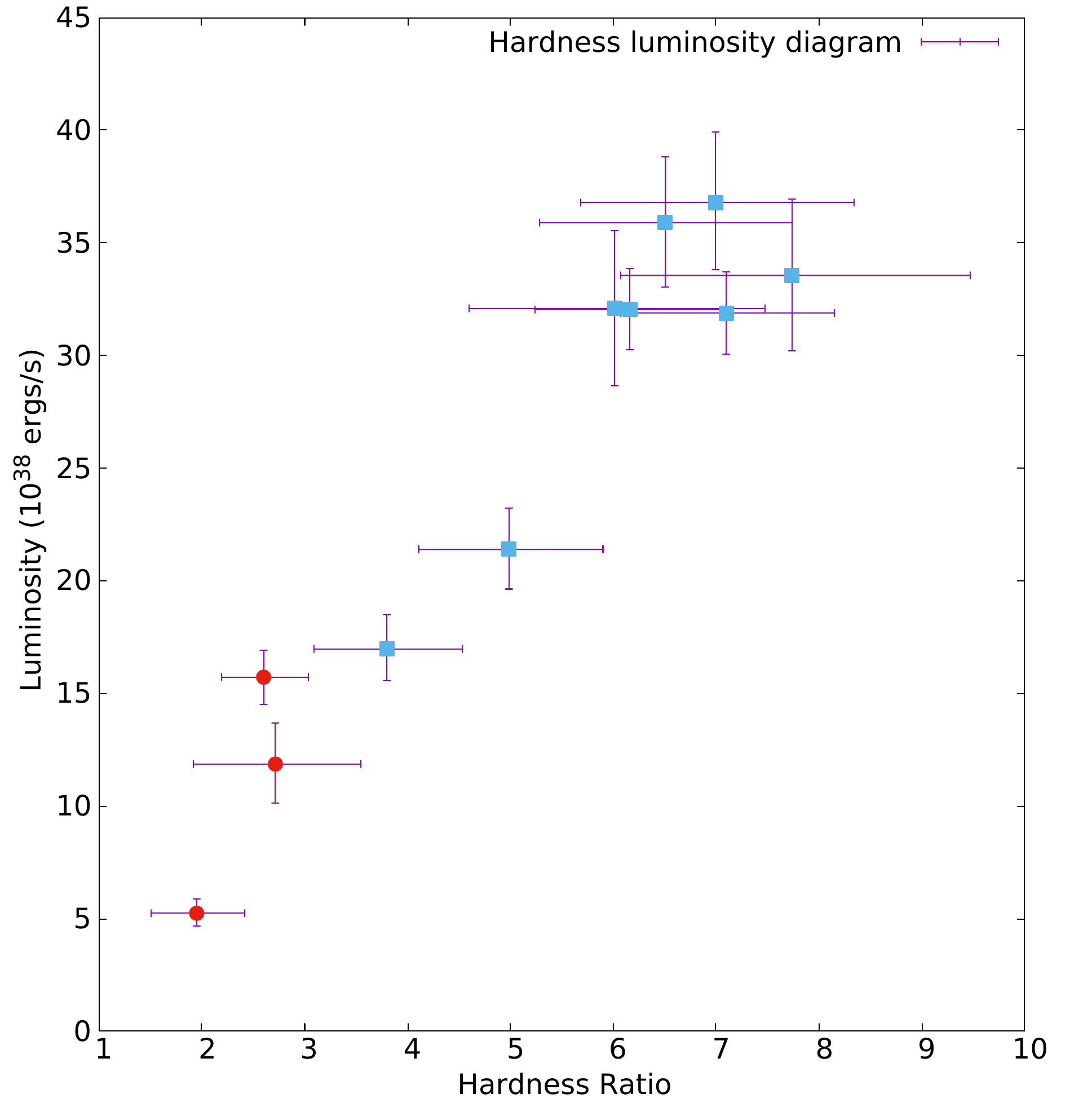}
    }

    \caption{Summary of spectral and timing analysis: (a) flaring-state fits, 
    (b) non-flaring-state spectra, (c) photon index $\Gamma$ evolution, and 
    (d) hardness–luminosity diagram.}
    \label{fig:spectra_summary}
\end{figure*}
\begin{table*}
\scalebox{0.90}{
\hspace{-3.5cm}
\centering
\setlength{\tabcolsep}{6pt}
\renewcommand{\arraystretch}{1.8}
\begin{tabular}{c|c|c|c|c|c|c|c}
\toprule

 Obs Ids & nH ($10^{22} cm^{-2})$ & $T_{0}$ = $T_{in}$ (keV) & $\tau$ & Norm(comptt)($10^{-5}$) & Norm(diskbb) & flux ($10^{-13}$ $ergs$ $cm^{-2}$$s^{-1}$) & $\chi^2/dof$   \\
\midrule
2057 &\multirow{11}{*}{$\prescript{\textcolor{red}{a}}{}{0.048}$} & \multirow{3}{*}{$\prescript{\textcolor{blue}{b}}{}{0.26^{+0.06}_{-0.06}}$} & $9.49^{+9.8}_{-3.4}$ & $0.55^{+0.39}_{-0.22}$& $0.29^{+0.94}_{-0.77}$ & 0.50 & 12/11\\
2058 &  &  & $10.74^{+3.9}_{-2.1}$ & $1.82^{+0.75}_{-0.49}$  & $0.72^{+0.94}_{-0.77}$ & 1.64& 36/38\\

0154350101 &  &  & $10.73^{+3.8}_{-2.2}$& $1.41^{+0.57}_{-0.40}$  & $0.52^{+0.94}_{-0.29}$&1.17& 27/26\\
\cline{3-3}

14801 & & \multirow{8}{*}{$\prescript{\textcolor{green}{c}}{}{0.61^{+0.15}_{-0.19}}$} & $5.61^{+3.0}_{-3.3}$&$4.02^{+1.5}_{-1.2}$& $0.03^{+0.03}_{-0.02}$ & 3.24& 12/11\\

16000 &  &  & $9.76^{+2.8}_{-1.7}$ & $3.19^{+1.2}_{-0.78}$ & $0.04^{+0.07}_{-0.02}$ & 3.70& 59/55\\

16001 &  &  & $7.80^{+5.6}_{-3.5}$ & $3.85^{+1.6}_{-1.2}$ & $0.04^{+0.05}_{-0.02}$& 3.90& 18/20\\

16484 &  &  & $7.90^{+4.2}_{-2.7}$ & $3.68^{+1.6}_{-1.1}$ & $0.04^{+0.06}_{-0.02}$& 3.80&20/21\\

16485 &  &  & $9.25^{+11}_{-4.5}$ & $3.22^{+1.3}_{-1.1}$ & $0.04^{+0.08}_{-0.02}$ & 3.63&13/10\\

16002 &  &  & $6.63^{+1.3}_{-1.6}$ & $3.60^{+1.2}_{-0.9}$ & $0.03^{+0.04}_{-0.01}$&  3.22&51/48\\
16003 & &  & $>9.5$ & $1.40^{+0.83}_{-0.51}$& $0.04^{+0.07}_{-0.02}$ & 2.34&29/34\\

0864270101 &  &  & $10.46^{+44}_{-3.1}$ & $1.04^{+0.98}_{-0.55}$ & $0.04^{+0.07}_{-0.02}$ &1.84&108/101\\

\bottomrule
\end{tabular}
}
\caption{Spectral fitting results of all observations using model \texttt{tbabs*(diskbb+comptt)}. Simultaneous fitting gives $\chi^2$/dof = 383/373.\hspace{0.2cm}\textcolor{red}{a :} nH fixed to the galactic value.\textcolor{blue}{\hspace{0.1cm}\textit{b}}, \textcolor{green}{\hspace{0.1cm}\textit{c}}: linked input photon temperature $T_{0}$ and inner disk temperature $T_{in}$ for observation with flaring state (Group A) and non-flaring state (Group B and Group C).}
\label{tab:diskbb+comptt_all_obs}
\end{table*}

\subsection{Spectral analysis}

We use Xspec v12.13.0 (\citealt{arnaud1996}) to perform the detailed spectral analysis of the ULX M74 X-1. To model the absorption effects caused by neutral absorbers, we have used \texttt{tbabs} with updated solar abundances (\citealt{Wilms2000}), with photoionization cross-section given by \citet{Verner1996}. We have used the $\chi^{2}$ minimization method for model spectral fitting and report the errors with a 90 percent confidence range unless stated otherwise.
\par

We begin the data analysis by plotting the unfolded spectra of all the observations listed in Table \ref{tab:obs_set} to look for any spectral variability by visual inspection. The unfolded spectra are produced using a \texttt{powerlaw} model with zero photon index, essentially a constant model. We plot the unfolded spectra of observations in flaring and non-flaring states separately to see any change in their spectral shape (Figure \ref{fig:unfolded_spectra}). All observations in the flaring state seem to have a similar spectral shape, with different flux values (Figure \ref{fig:unfolded_spectra_flaring}). 
Here, variability is seen in both soft (0.3 - 1.0 keV) and hard (1.0 - 10.0 keV) energy bands. On the other hand, the non-flaring states have curved spectra that seem to overlap below 1 keV and variable above 1 keV  (Figure \ref{fig:unfolded_spectra_non_flaring}). Out of the eight observations in the non-flaring state, the first six observations (Obs IDs - 14801, 16000, 16001, 16484, 16485, and 16002) show overlapping spectra in the hard X-ray band ( $>$1.0 keV). The remaining two observations (Obs. IDs: 16003 and 0864270101) also have overlapping spectra in the hard X-ray band ( $>$1.0 keV), but the hard X-ray flux of the last two observations ((1.52 - 2.03) $\times$ $10^{-13}$ $ergs$ $cm^{-2}$$s^{-1}$) is less than the first six observations ((2.89 - 3.51) $\times$ $10^{-13}$ $ergs$ $cm^{-2}$$s^{-1}$). We begin the spectral analysis with the basic absorbed \texttt{powerlaw} model to see the variation of nH.  While there are fluctuations in the nH values across different epochs, the values overlap within the error bars.
Moreover, because of the spectral difference between flaring and non-flaring states, we analyze them into separate groups.
For the observations in the flaring state, linking nH provides statistically similar results (${\chi}^2/dof$= $84/74$ to ${\chi}^2/dof$ = $86/76$). Here, nH is not constrained $<$ 0.05, so we further freeze nH to the galactic value of 0.048 ($10^{22}$ $atoms/cm^2$), which gives $\chi^2/dof$=88/77. It provides an acceptably good fit with excess around 1 keV and some marginal excess at hard energies (above 3 keV) in the residuals, as can be seen in Figure \ref{fig:flaring_po_diskbb_gauss}. 
We add the \texttt{diskbb} component in addition to the power law to see if it provides a better fit. It improves the fit with ${\chi}^2/dof$= $71/71$. The \texttt{diskbb} component takes care of the low-energy part in the spectra, whereas the power law is responsible for explaining the hard X-ray part. We see that the inner disk temperature T$_{in}$ and $\Gamma$ values for each spectrum are within error bars, so we further link the T$_{in}$ and $\Gamma$ values between these observations, which gives ${\chi}^2/dof$= $73/75$. (Table \ref{tab:flaring_obs_fitting}). Similar inner disk temperature $T_{in}$ and $\Gamma$ values between the observations in the flaring state indicate that the spectra of the flaring state have identical shapes with varying soft and hard X-ray component flux. Due to the spectral similarity among the observations in the flaring state, we group all the observations in the flaring state in Group A. We also observe excess residue around 1 keV in the spectra of the flaring state, which was not reported in the previous studies. Such a feature has been observed in several other ULXs, which is believed to arise due to the blend of atomic absorption and emission lines formed by the interaction of hard photons with the wind. These signatures of wind indicate the presence of super-Eddington accretion in the system. We add a \texttt{Gaussian} component to model this 1 keV feature, where we link the Gaussian line energy $E_{l}$ and line width $\sigma$ between the observations. This gives $E_{line}$ = $0.96^{+0.05}_{-0.11}$ keV with $\sigma$ = $0.11^{+0.13}_{-0.06}$ keV. Adding \texttt{Gaussian} in \texttt{diskbb+powerlaw} model improves the statistics with ${\chi}^2/dof$= $55/68$. There is a degeneracy between the \texttt{Gaussian} and \texttt{diskbb} components in the low energy part, hence the addition of the Gaussian component lowers the inner disk temperature to $T_{in}$=$0.15^{+0.12}_{-0.07}$ keV but has overlapping values within errorbars with the disk temperature when the \texttt{Gaussian} component is not included ($Tin=0.28^{+0.04}_{-0.04}$ keV).

\par
For observations in the non-flaring state, linking the nH for the \texttt{powerlaw} model changes the statistics from ${\chi}^2/dof$= $322/301$ to ${\chi}^2/dof$= $333/308$, which is statistically similar. Therefore, nH is linked for further analysis. The power law model provides a sufficiently good fit to the data, with no requirement of additional \texttt{diskbb} component, unlike observations in the flaring state. There is no hard excess or residue around 1 keV in the spectra of the non-flaring state (Figure \ref{fig:non_flaring_spectra_res}). $\Gamma$ of the first six observations are consistent ($\Gamma$ $\sim$ 1.6) within the error bars, except for the last two observations (Obs. IDs: 16003 and 0864270101; Figure \ref{fig:gamma_evolution_non_flaring}). The last two observations are relatively softer and have flux values lower than the first six observations. The first six observations of the non-flaring group, which have similar parameter values, are further grouped into Group B. The last two observations, which also have similar parameter values, are grouped into Group C. Group B contains Obs. IDs - 14801, 16000, 16001, 16484, 16485, and 16002, and Group C contains Obs. IDs 16003 and 0864270101. To see the spectral difference between the groups, we plot one representative spectrum from each group (Figure \ref{fig:unfolded_spectra_groups}). Here, we can see different spectral shapes corresponding to the three groups.
To check for the presence of any spectral cutoff, a characteristic feature found in ULXs, we also employed the cutoff power-law model (\texttt{cutoffpl}) to fit the spectra. For observations in a non-flaring state, we link the nH values between the observations and separately link the e-folding energy $E_{fold}$ in Group B and Group C. It gives a sufficiently good fit with ${\chi}^2/dof$= $318/306$, significantly improving the power law model with $\Delta$$\chi^2$ = 15 for 2 d.o.f. We see that gamma values are within error bars for Group B and Group C separately, which we linked in the further analysis. We get e-folding energy of $E_{fold}$ = $3.9^{+2.4}_{-1.1}$ keV and $E_{fold}$ $>$ 4.1 keV for Group B and Group C, respectively. Since the e-folding energies for both groups overlap, we link them, resulting in $E_{fold} = 4.6^{+3.4}_{-1.4}$ keV (Table \ref{tab:non_flaring_cutoffpl}), with ${\chi}^2/dof = 327/313$. Fitting the spectra of the flaring state with the \texttt{cutoffpl} model results in an unconstrained e-folding energy. This is also evident in the spectral shape of the observations in the flaring state, which increases sharply after $\sim$ 3 keV. We further freeze the e-folding energy to $E_{fold}$ = 4.6 keV for the flaring state, which we obtain from the fitting of observations with the non-flaring state, to see how the statistics change. It gives a poor fit with ${\chi}^2/dof$ = $109/77$ and excess in the residuals.

\subsubsection{\texttt{diskbb+comptt}}
The \texttt{diskbb+powerlaw} provides a good fit for the flaring state. However, in this section, we discuss the use of a more physically motivated model, \texttt{diskbb+comptt}, where \texttt{powerlaw} in the previous model is replaced by \texttt{comptt}. Although \texttt{diskbb} is not statistically required in the non-flaring state, we employ this model for both flaring and non-flaring states to compare the relative contributions of the soft and hard components in both these states. 
\par
The \texttt{comptt} component considers the Comptonization of soft photons by the optically thick corona, where the soft photons are provided by the accretion disk. This interpretation is generally used in the case of sub-Eddington accretion in GBHBs. An alternate interpretation of this model has been proposed where the soft photons are attributed to the massive optically thick out-flowing wind instead of the accretion disk, with the hard photons coming from the inner regions of the disk or corona \citep{kajava2009}. This wind could launch from the inner regions of the disk if the accretion flow is super-Eddington. The detection of a 1keV Gaussian feature in the flaring state of M74 X-1 could indicate a super-Eddington accretion flow with an outflowing wind in the system. Therefore, we adopt the second interpretation of \texttt{comptt} in our work.
\begin{figure*}
    \centering

    \subfigure[Variation of normalization of the cool \texttt{diskbb} component (Norm(cool)) for observations in the non-flaring state, using model \texttt{tbabs*(diskbb+diskbb)} with linked $N_{\rm H}$, $T_{\rm in}$(cool), and $T_{\rm in}$(hot). 
    \label{fig:norm_cool_evolution}]{
        \includegraphics[width=0.45\linewidth]{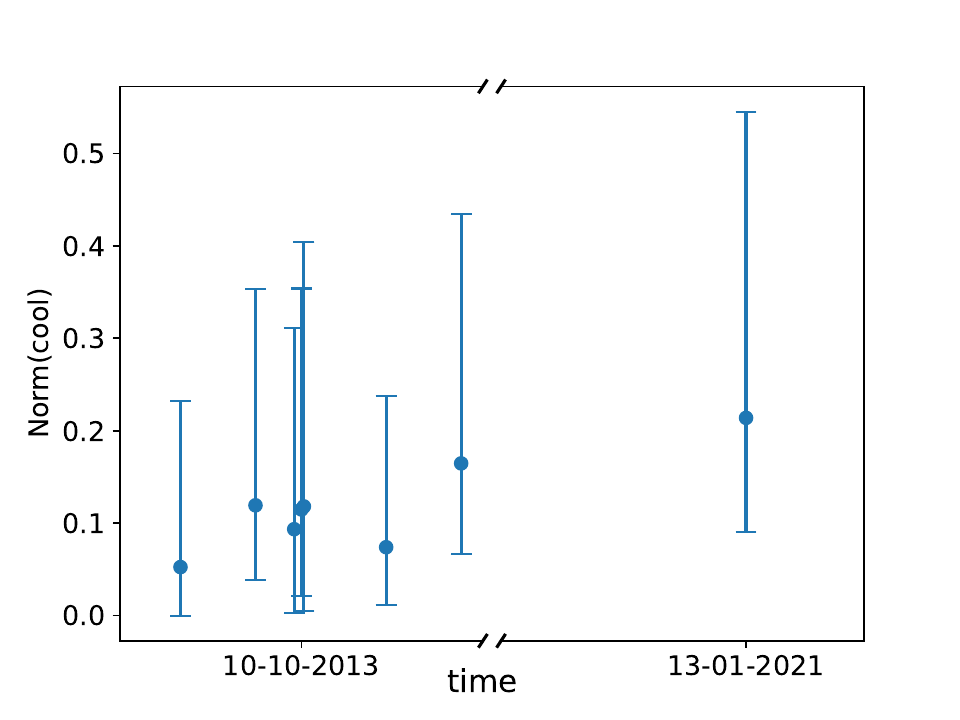}
    }
    \hfill
    \subfigure[Variation of normalization of the hot \texttt{diskbb} component (Norm(hot)) for observations in the non-flaring state, using model \texttt{tbabs*(diskbb+diskbb)} with linked $N_{\rm H}$, $T_{\rm in}$(cool), and $T_{\rm in}$(hot). 
    \label{fig:norm_hot_evolution}]{
        \includegraphics[width=0.45\linewidth]{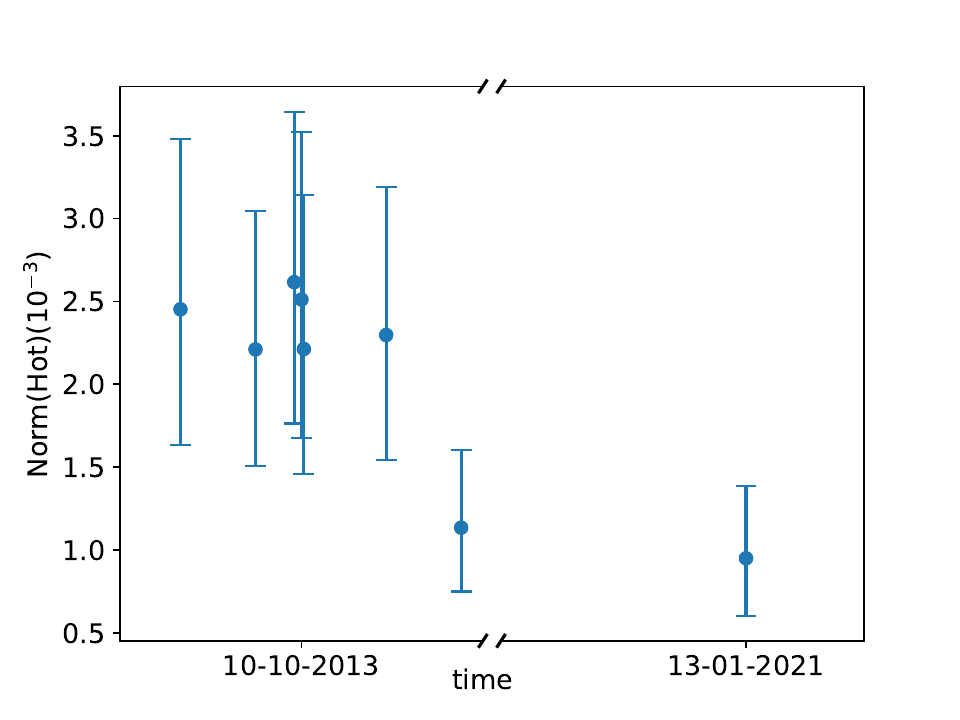}
    }

    \caption{Evolution of the \texttt{diskbb} component normalizations in non-flaring state observations: 
    (a) Norm(cool) and (b) Norm(hot).}
    \label{fig:Norm_cool_hot_evolution}
\end{figure*}
\par
 In the \texttt{diskbb+comptt} model, we initially allow the input photon temperature $T_{0}$ to vary independently for Groups A, B, and C. Since Groups B and C show similar $T_{0}$ values, we link them for further analysis. Similarly, we initially allow the inner disk temperature $T_{in}$ to vary independently for these groups.  However, because Groups B and C have overlapping spectra below 1 keV, we link $T_{in}$ of the soft component \texttt{diskbb} between them. The plasma temperature we obtain is not well-constrained, with the best-fit value around $\sim$ 2.0 keV. Therefore, we freeze it at $T_e \sim 2.0$ keV, which is consistent with the values generally observed in ULXs (\citealt{pintore2014}). Further, we set the input photon temperature equal to the inner disk temperature, $T_{0}$ = $T_{in}$. This provides us with a good fit with ${\chi}^2/dof$= $383/373$. For Group A, the temperature comes out to be $T_{0}$ = $T_{in}$ = $0.26^{+0.06}_{-0.06}$ keV, whereas for Group B and C,  $T_{0}$ = $T_{in}$ = $0.61^{+0.15}_{-0.18}$ keV (Table \ref{tab:diskbb+comptt_all_obs}). We get very high optical depth $\tau$ $\sim$ 8.0 for plasma temperature of $T_e$ = 2.0 keV for both flaring and non-flaring states. This state is different from the state we generally find in GBHBs, where we get hot and optically thin corona $\tau$ $\sim$ 1 (\citealt{gladstone2009}; \citealt{pintore2014}). We see that the optical depth is consistent within error bars for both flaring and non-flaring states. We confirmed this by linking the optical depth of all observations, which changes the statistics from ${\chi}^2/dof$= $383/373$ to ${\chi}^2/dof$= $401/383$, practically providing the same fit statistics. This gives us a common optical depth of $\tau$ = $8.6^{+0.76}_{-0.53}$.
\par
We plot the Hardness luminosity diagram (Figure \ref{fig:Hardness_luminosity_diagram}) for both flaring (red circle) and non-flaring states (blue square). Here, hardness is defined as the flux ratio between (1.0 - 10.0 keV) and (0.3 - 1.0 keV), which is calculated using \textit{cflux} convolution model. The luminosity is unabsorbed, calculated for the energy range (0.3 - 10 keV). We find that the hardness ratio is lower in the flaring state compared to the non-flaring state, indicating a higher fraction of soft X-ray flux during flares. We see a trend of increasing hardness with luminosity, which can be explained if the system is being viewed at low inclination. With the increase in accretion rate, the wind opening angle decreases, which causes more hard photons to be geometrically beamed into our line of sight \citep{middleton2015(a)}. However, the presence of a 1 keV feature in low-flux observations suggests that the system might instead be viewed at a moderate inclination, where the wind intersects our line of sight and causes overall flux absorption. Therefore, within the limitations of current data, it is hard to obtain a clear geometrical picture of the source, as the two results point to different scenarios.
 
\subsubsection{\texttt{diskpbb}}
Due to the curved spectral shape observed in the non-flaring state, the \texttt{diskpbb} model also provides a good fit to the data. It is a multi-temperature disk black body where temperature varies as a function of r, where r is the radial distance along the accretion disk as T $\propto$ $r^{-p}$ (\citealt{mineshige1994}). For p = 0.75, it is a standard thin accretion disk (\citealt{shakura1973}) accreting at sub-Eddington rate; for p $<$ 0.75, the disk is affected by advection; for p = 0.5, it is referred to as the slim disk \citep{abramowicz1988}, signifying the presence of super-Eddington accretion in the system. Fitting the spectra of the flaring state with \texttt{diskpbb} gives unphysically high inner disk temperatures $T_{in}$ $>$ 5 keV. For the spectra of non-flaring state, we did the analysis of Group B and Group C separately and linked the inner disk temperatures within each Group separately. It provides a sufficiently good fit to the data, with p values within error bars within each group. We further link the p-values in these Groups, which changes the statistics from $\chi^2$/dof = 315/306 to $\chi^2$/dof = 321/312, which is not statistically different. Inner disk temperature for group B comes out to be $T_{in}$ = $2.07^{+0.55}_{-0.31}$ keV with p = $0.60^{+0.04}_{-0.02}$ whereas for the group C it is $T_{in}$ = $3.5^{+1.5}_{-1.5}$ keV with p $<$ 0.53. The p values in both groups are lower than p = 0.75 of the standard thin accretion disk, suggesting the advection-dominated flow in the system. Although Group C has a higher best fit value of $T_{in}$ than Group B, these values are within the error bars. We confirmed this by linking the temperature of the two groups together. This gives us $\chi^2$/dof = 324/313 with a common temperature of $T_{in}$ = $2.12^{+0.64}_{-0.32}$ keV for both Groups. For a common temperature, group B has a higher p-value of p = $0.61^{+0.04}_{-0.03}$ with a high flux value as compared to group C with p = $0.52^{+0.04}_{-0.02}$ and a low flux value.

\subsubsection{\texttt{diskbb+diskbb}}
Because of the curved spectral shape of the observations in the non-flaring state and their non-variability below 1 keV, we also fit the spectra with a dual thermal model \texttt{diskbb+diskbb}. Here we link the inner disk temperature $T_{in}$ of the cool \texttt{diskbb} component among the observations, which takes into account the spectral invariability below 1 keV, keeping the hot \texttt{diskbb} component to vary, which is useful in producing the spectral curvature in the spectra. The hot \texttt{diskbb} represents the emission from the inner regions of the disk or corona, and cool \texttt{diskbb} represents the emission from the outer disk or the massive optically thick outflows. The motivation for using this model for observations in the non-flaring state further comes from the fact that the best XMM observation (Obs. ID - 0864270101) with the highest counting statistics provides a good fit for this model with all the parameters well constrained. 
\par
Fitting the spectra of the flaring state with this model gives very high hot temperature $T_{in}$(hot) $\sim$~5.0 keV, and very low normalization values $\sim$ ($10^{-6}$ - $10^{-4}$). Fitting the non-flaring spectra with this model, we link the $n_{H}$ and the cool temperature $T_{in}(cool)$ of all observations as the non-flaring spectra overlap below 1 keV and separately link the hot temperature within Groups B and C. It provides an acceptable fit to the data with $\chi^2$/dof = 310/305. We obtain a cool temperature of $T_{in}(cool)$ = $0.41^{+0.11}_{-0.08}$ keV, and a hot temperature of $T_{in}(hot)$ = $1.64^{+0.18}_{-0.14}$ keV for Group B and $T_{in}(hot)$ = $1.88^{+0.65}_{-0.33}$ keV for Group C, respectively. Most of these $T_{in}(hot)$ values are consistent within error bars, so we further link the $T_{in}(hot)$ between Groups B and C. This does not change the statistics and gives a common cool temperature $T_{in}(cool)$ = $0.38^{+0.08}_{-0.06}$ keV, hot temperature $T_{in}(hot)$ = $1.67^{+0.18}_{-0.13}$ keV along with $n_{H}$=$0.12^{+0.05}_{-0.04}$ ($10^{22}cm^{-2})$ for both Group B and C, with $\chi^2$/dof = 311/306. Plotting the normalizations of cool temperature component ($Norm(cool)$), we see the values are consistent within error bars (Figure \ref{fig:norm_cool_evolution}), which explains the non-variability in the spectra in the soft X-ray part (below 1 keV). Plotting the normalizations of hot temperature component ($Norm(hot)$), we see that the 
 values are within error bars for Groups B (first six points of figure \ref{fig:norm_hot_evolution}) as well as for Group C (last two points of figure \ref{fig:norm_hot_evolution}) separately. This indicates that the spectral shapes and flux values within Groups B and C are similar, with Group B having higher flux values than Group C.
 
\subsubsection{Significance of 1 keV feature in the spectra of flaring and non-flaring states}
We detect a 1 keV Gaussian feature in the spectra of the flaring state observations. However, this feature is not significant during the observations of the non-flaring state. To compare the significance of this feature between flaring and non-flaring states, we employ \texttt{simftest} in the \texttt{xspec}, which generates simulated datasets and uses these datasets to calculate the F-test probability for adding a model component. We use this test to estimate the significance of the Gaussian component in the model \texttt{tbabs*(po+gauss)} for both flaring and non-flaring states. We choose \texttt{powerlaw} model as the continuum because it provides sufficiently good statistics for both flaring and non-flaring states with constrained parameters. We fit this model separately for flaring and non-flaring states, linking the line energy $E_{l}$ and line width $\sigma$ between the observations. For flaring spectra, we get a line energy of $E_{l}$ $\sim$ 0.96 keV with a line width of $\sim$ 0.1 keV, which is similar to the values generally observed in the ULXs (\citealt{earnshaw2017}; \citealt{middleton2015(b)}; \citealt{ghosh2023}; \citealt{ghosh2022}; \citealt{pintore2021}; \citealt{Earnshaw2019}). The addition of the Gaussian component improves the statistics from $\chi^2$/dof = 88/77 to $\chi^2$/dof = 68/73 in the flaring state, which is an improvement of $\Delta$$\chi^2$ = 20 for 4 d.o.f. For the non-flaring state, the line energy and line width are not constrained for most of the observations, so we freeze the line energy $E_{l}$ = 0.96 keV and line width $\sigma$ = 0.1 keV and see the normalization values of the Gaussian component in these observations. We see that only for observations with Obs. IDs - 16000, 16001, 16484, 16485, and 16003 have normalization values that are not zero. We further do the simultaneous fitting of these observations where we link the $E_{l}$ and $\sigma$ between them. This changes the statistics from $\chi^2$/dof = 142/145 to $\chi^2$/dof = 135/139, which is not statistically different. 
\par
We do the \texttt{simftest} for both the analysis (for flaring and non-flaring states) for 10000 iterations, and we got $99.84$ \%  confidence interval, which is a $\sim$ 3$\sigma$ significance result for the flaring state. For a non-flaring state, this comes out to be $78.3$ \%, which is a $\sim$ 1.2 $\sigma$ result. This strongly suggests the prominent presence of a 1 keV line feature in the flaring state, as compared to the non-flaring state.

\section{Discussion}
\label{sec:discussions}
\subsection{Spectral difference between flaring and non-flaring states}

We analyze 11 XMM-Newton and Chandra observations of M74 X-1 between 2001 and 2021 and study its long-term spectral and timing properties. Initial observations (Obs. IDs: 2057, 2058, and 0154350101) were studied by \citealt{LIU2005} and \citealt{Krauss2005}, which revealed flaring behavior in the source. For completeness, we reanalyze these observations in the flaring state and more recent observations of the source that did not show any flaring events. The flaring states with lower average flux show two-component-like spectra, which include a soft thermal component and a hard power-law component. In contrast, non-flaring states have higher flux and curved single-component spectra. Comparing the general spectral variability of ULXs with M74 X-1, we see that similar spectral evolution has been observed in other ULXs also, such as  NGC 1313 X-1 (\citealt{dewangan2010}), NGC 1313 X-2 (\citealt{robba2021}), Ho \Romannum{9} X-1 (\citealt{luangtip2016}; \citealt{walton2017}; \citealt{jithesh2017}). 
 \par
 For the flaring state, we identify the best fit model as \texttt{ tbabs * (diskbb + po)} with $\Gamma=1.3^{+0.30}_{-0.35}$ and $Tin=0.28^{+0.04}_{-0.04}$ keV.
 We also detect a 1 keV line feature with $E_{line}=0.96^{+0.05}_{-0.11}$ keV and $\sigma=0.11^{+0.13}_{-0.06}$ keV in the spectra of the flaring state, which is not significant in the non-flaring state. Such emission feature has been reported in several other ULXs like NGC 6946 X-1, NGC 4395 X-1, NGC 1313 X-1, NGC 5408 X-1, NGC 55 X-1, Ho \Romannum{9} X-1, Ho \Romannum{2} X-1 (\citealt{middleton2015(b)};  \citealt{ghosh2022}; \citealt{ghosh2023}; \citealt{pinto2016,pinto2017}). These residues suggest the presence of outflows in the system, which further implies the presence of super-Eddington accretion in the system. 
 \par
 From the hardness-luminosity diagram, we see that flaring states (showing the 1 keV feature) have a higher fraction of soft X-ray flux than non-flaring states. This has been reported for several ULXs where the 1 keV feature was most prominent in the softer spectra \citep{kosec2021}. The variation of 1 keV feature from soft (flaring) to hard (non-flaring) spectra can be explained using the wind model in super-Eddington accretion (\citealt{poutanen2007}; \citealt{middleton2015(b)}). If our line of sight passes through the wind, we observe emission and absorption features in the spectra that are formed by the interactions of hard photons with the wind. We also receive a high fraction of the soft X-ray flux due to the downscattering of hard photons from the wind. Therefore, the presence of the 1 keV Gaussian feature and a higher fraction of soft X-ray flux in the observations with flaring state suggest that the wind is in our line of sight during the flaring state, whereas it is out of our line of sight during non-flaring states. The presence of wind features in low flux observations suggests that the system is being viewed at moderate inclinations where the wind is crossing our line of sight. However, the observed increase in hardness with luminosity suggests that the system is being viewed at a low inclination where the central regions of the accretion disk are not completely obscured by the outflowing wind. At low inclination, an increase in the accretion rate causes the wind funnel to narrow, leading to more hard photons being geometrically beamed into our line of sight (\citealt{king2009}). This results in the observed trend of an increase in hardness with luminosity. Similar behavior was observed in ULXP NGC 1313 X-2, where traces of wind were found in low flux observations and the trend of increasing hardness with luminosity \citep{middleton2015(a)}. Recently, a harder when brighter effect along with the 1 keV residue feature in the spectra was also observed in one potential ULX Pulsar candidate, NGC 4559 X-7 \citep{pintore2025}. To further investigate the presence of the wind feature during the flaring state at low and high count rates, we divide the spectrum of Obs ID: 2058 in the flaring state based on the count rate. We select Obs ID: 2058 for this analysis since flaring is the most prominent and regular in this observation, and also has sufficient statistics for such an analysis. We extract the spectrum with a count rate $>$ 0.03 count/s and call it the bright spectrum. Similarly, we extract the spectrum with a count rate $<$ 0.03 count/s and call it the faint spectrum. We model both spectra with a power-law continuum and add a Gaussian component to account for the 1 keV residue due to the wind. We do the \texttt{simftest} on the Gaussian component for 10000 iterations to check the significance of this Gaussian component. We find a 99.7\% confidence interval for the faint spectrum, which is a $3\sigma$ significance result, whereas, for the bright spectrum, we get an 85.8\% confidence interval, which is equivalent to the 1.44$\sigma$ result. This shows the wind feature to be more prominent in the faint spectra as compared to the bright spectra. This can occur if the wind crosses our line of sight, resulting in significant photon absorption and causing a low flux state with enhanced wind features. We tried a similar analysis with other observations in the flaring state, but the statistics were not sufficient for such a division of spectra.\\
 A 1 keV wind feature along with a QPO has also recently been observed in one observation of the ULXP M51 ULX-7 \citep{imbrogno2024}, where a similar super-Eddington accreting wind model was used to explain the QPO. Here, the puffed-up accretion disk, formed due to super-Eddington accretion, generates an outflowing wind with quasi-periodic recurrence, resulting in quasi-periodic variations in the light curve.  We have discussed more about the presence of wind and its association with QPO in Section 4.3.
 
\subsection{Mass and Inner Radius estimates of the compact object and accretion disk}

 Using \texttt{diskbb+diskbb} model, we get cool temperature of $T_{in}(cool)$ = $0.38^{+0.08}_{-0.06}$ keV whereas the hot temperature is $T_{in}(hot)$ = $1.67^{+0.18}_{-0.13}$ keV for observations in non-flaring state. Plotting the normalizations of the two \texttt{diskbb} components (Figure \ref{fig:Norm_cool_hot_evolution}), we see that $Norm(cool)$ stays within the error bars (Figure \ref{fig:norm_cool_evolution}), signifying no variability below $\sim$1 keV. The normalization of hot \texttt{diskbb} $Norm(hot)$ (Figure \ref{fig:norm_hot_evolution}) is within error bars for Groups B and C separately, with Group B having most of the values higher than that of Group C. We calculate the inner radius of the accretion disk from the normalization factor of hot \text{diskbb} component using the formula $R_{in}$ = $\xi$$\kappa^2$$\sqrt{(Norm*D_{10}^{2})/cos(i)}$, where $D_{10}$ is the distance to the source in units of 10 kpc, $i$ is the angle of inclination of the disk with our line of sight, $\xi$ is the geometric factor, $\kappa$ is the color correction factor \citep{kubota1998}. For \texttt{diskbb}, we take $\xi$ = 0.412 and $\kappa$ = 1.7 \citep{kubota1998}.  
 As the $Norm(hot)$ is different for Groups B and C, we calculate hot  \texttt{diskbb} inner radii for both groups separately. We get $R_ {in,B}(hot)$ =$54^{+10}_{-10}$ $(cosi)^{-1/2}$ km and $R_{in, C}(hot)$ = $36^{+7}_{-7}$ $(cosi)^{-1/2}$ km respectively. Assuming a moderate inclination angle of $i$ $\sim$ 60 $\degree$, inner radius comes out to be $R_{in, B}(hot)$ = $77^{+14}_{-14}$ km and $R_{in, C}(hot)$ = $50^{+11}_{-9}$ km. To calculate the compact object's mass from the inner radius of the accretion disk, we use the average radius $R_{in, avg}(hot)$ of Groups B and C, which comes out to be $R_{in, avg}(hot)$ = $64^{+13}_{-12}$ km. Assuming that for the black hole compact object, the inner radius of the accretion disk touches the Inner Stable Circular Orbit (ISCO), we get the mass of the compact object as M = $7.1^{+1.4}_{-1.3}$ M$_{\odot}$. This comes under the stellar Mass Black Hole (sMBH) regime and signifies super-Eddington accretion in the system. Using the steps highlighted in Section 4.4 of \citealt{ghosh2023}, we calculate the mass accretion rate in the unit of Eddington accretion rate, i.e., $\dot{m}_{0} = \dot{M}_{0}/\dot{M}_{Edd}$ and the spherization radius $R_{sph}$ for both Groups B and C. Taking the mean luminosity from each group, $L_{B}$ = 3.48 $\times$ $10^{39}$ ergs/s and $L_{C}$ = 1.94 $\times$ $10^{39}$ ergs/s, and taking $m_1$ = 7.1, we obtain $\dot{m}_{0}$ $\approx$ 9 and $\dot{m}_{0}$ $\approx$ 7 for Groups B and C, respectively. The spherization radius comes out to be $R_{sph, B}$ = $1296^{+264}_{-243}$ km and  $R_{sph, C}$ = $1008^{+205}_{-189}$ km.
 The inner radius of the hot \texttt{diskbb} component is smaller than the spherization radius of both groups $R_{in (B, C)}(hot)$ $<$ $R_{sph(B, C)}$, indicating that this component is likely reproducing the super-Eddington inner accretion flow within the spherization radius $R_{sph}$.

\subsection{Quasi-periodic modulation, its explanation and association with wind}

We detect a 1 keV wind feature in the spectra of the observations in the flaring state. These same observations have previously been reported to exhibit quasi-periodicities by \cite{Krauss2005} and \cite{LIU2005}. It is similar to the heartbeat oscillations shown by 4XMM J111816.0–324910 \citep{motta2020} and quasi-periodic modulations shown by 4XMM J140314.2+541806 \citep{urquhart2022}. Similar modulation with the traces of wind in the same observation has also been recently reported in a ULXP M51 ULX-7 \citep{imbrogno2024}. All these sources have similar timescales of modulation, giving rise to the mHz QPOs. Previous studies have proposed various models to explain these modulations, such as the limit cycle instability driven by the Lightman-Eardley radiation pressure instability for 4XMM J111816.0–324910; Lense-Thirring precession of an outflow from the inner regions of the disk \citep{middleton2019} for 4XMM J140314.2+541806 and M51 ULX-7. To understand the quasi-periodic modulation in M74 X-1, we divide the spectrum of the observation with flaring state, Obs Id: 2058, (showing highest modulation in the lightcurve), on the basis of the count rate: Bright spectra (count rate $>$ 0.03 counts/s) and Faint spectra (Count rate $<$ 0.03 counts/s) and fit it with the model \texttt{diskbb+diskbb} to see the evolution of spectral parameters of the hard and soft spectral components with the count rate. We found that the spectral parameters do not change with the count rate, which is in contrast with the source 4XMM J111816.0–324910, where a clear evolution in the spectral parameters has been observed in the different phases of the modulation. Also, modulation in 4XMM J111816.0–324910 is more regular and stable, similar to the $\rho$ class variability found in GRS 1915+105 \citep{belloni2000}, whereas the variability in M74 X-1 is quite irregular. All these observations suggest that modulation in M74 X-1 is different from that in 4XMM J111816.0–324910, and thus, limit cycle instability may not be the possible explanation for M74 X-1 QPO. Compared to other sources, the modulation properties of M74 X-1 closely resemble those of M51 ULX-7, where similar irregularity in the flares with no change in the spectral parameters with count rate has been observed \citep{imbrogno2024}. Another similarity between M74 X-1 and M51 ULX-7 is the presence of a 1 keV wind feature in the same observations where a QPO is reported, which suggests that Lense-Thirring precession (used to explain QPOs in M51 ULX-7) could be responsible for the quasi-periodicity in M74 X-1 as well. Moreover, the occurrence of QPOs in certain observations and their absence in others further suggests the potential presence of precession in the system. As \citep{middleton2019} pointed out, the strength of the QPO feature is influenced by the inclination angle of the precessing system. If the inclination angle is small (with a small precessing cone), QPO features tend to be weak because the outflowing wind does not obscure our line of sight. However, as the inclination angle increases, the QPO features become more pronounced. This occurs because the line of sight alternates between pointing directly at the central accretion disk and passing through the wind, leading to modulation in the light curve. This might explain why QPO features are present in only a few observations while being absent in other observations. However, unlike M51 ULX-7, which exhibits a $\sim$39-day super-orbital periodicity \citep{vasilopoulos2020a}, attributed to the precessing wind period ($P_{\text{wind}}$) in the context of Lense-Thirring precession \citep{middleton2019}, no such periodicity has been reported for M74 X-1. Therefore, this explanation remains unconfirmed and further requires long-term monitoring of the source to confirm this.

\section{Conclusions}
\label{sec:conclusions}
 In this paper, we report the long-term spectral and timing evolution of Ultraluminous X-ray source M74 X-1 using 11 archival XMM-Newton and Chandra observations. We see that some observations show flaring activity in them, whereas in others, it is absent. Observations in the flaring state have two-component spectra with low flux values, whereas the non-flaring state has single-component spectra with high flux values. This spectral evolution is very similar to the spectral evolution of NGC 1313 X-1 (\citealt{dewangan2010}), NGC 1313 X-2 (\citealt{robba2021}), Ho \Romannum{9} X-1 (\citealt{luangtip2016}; \citealt{walton2017}; \citealt{jithesh2017}). Spectra of both flaring and non-flaring states are well described by models such as \texttt{powerlaw}, \texttt{diskbb+powerlaw}, and the Comptonizing corona model \texttt{diskbb+comptt} with low plasma temperature and high optical depth. Because of the curved spectral shape of observations in the non-flaring state, they are also well described by other models such as \texttt{cutoffpl}, \texttt{diskbb+diskbb}, \texttt{diskpbb}. The flaring state has softer spectra compared to the non-flaring state and displays a 1 keV Gaussian feature in the spectra, which is not significant in the non-flaring state. This feature has been reported in several ULXs and is more prominent in the softer spectra, aligning with current findings. This feature could indicate the presence of powerful outflows, which further indicates the presence of super-Eddington accretion in the system. QPOs were reported for this source by \cite{Krauss2005} and \cite{avdan2024} 
 which can be explained by the presence of wind into our line of sight with quasi-periodic occurrence or the Lense-Thirring precession of an outflow from an inner region of the disk, similar to the ULXP M51 ULX-7. \cite{Krauss2005} suggested that M74 X-1 is an Intermediate Mass Black Hole Candidate (IMBH) on the basis of low accretion disk temperature. In our analysis, we estimate the mass of the compact object from the hot \texttt{diskbb} component in \texttt{diskbb+diskbb}, assuming that the inner accretion disk radius truncates at the ISCO, which comes out to be $M = 7.1^{+1.4}_{-1.3}$ M$\odot$. This comes under the stellar Mass Black Hole (sMBH) regime, with super-Eddington accretion happening in the system. Further, the similarities between the M74 X-1 and the known ULXP, such that NGC 1313 X-2 and M51 ULX-7, suggest that it could also be a potential neutron star candidate.

\section{Data availability}
 This paper utilizes XMM-Newton and Chandra archival data available at the High Energy Astrophysics Science Archive Research Center (HEASARC) (accessible at \url{https://heasarc.gsfc.nasa.gov/cgi-bin/W3Browse/w3browse.pl}). The Chandra data sets are obtained by the Chandra X-ray Observatory, contained in~\dataset[DOI: 10.25574/cdc.347]{https://doi.org/10.25574/cdc.347}.
\bibliography{sample631}{}
\bibliographystyle{aasjournal}
\end{document}